\documentclass[superscriptaddress,twocolumn,showpacs,
amssymb,amsmath,nobibnotes,aps,prd,
nofootinbib]{revtex4-1}
\pdfoutput=1
\usepackage{graphicx,subfigure,bm,color,psfrag,hyperref}
\usepackage{amsfonts}
\usepackage{lipsum}
\usepackage{mathtools}
\usepackage{verbatim}
\usepackage[normalem]{ulem}
\usepackage[dvipsnames]{xcolor}
\hypersetup{colorlinks,linkcolor={blue},citecolor={red},urlcolor={blue}}

\begin{document}

\title{Evidence of dynamical dark energy in a non-flat universe: current and future observations}

\author{Mehdi Rezaei}
\email{rezaei@irimo.ir}
\affiliation{Iran Meteorological Organization, Hamedan Research Center for Applied Meteorology, Hamedan, Iran}

\author{Supriya Pan}
\email{supriya.maths@presiuniv.ac.in}
\affiliation{Department of Mathematics, Presidency University, 86/1 College Street, Kolkata 700073, India}
\affiliation{Institute of Systems Science, Durban University of Technology, PO Box 1334, Durban 4000, Republic of South Africa}

\author{Weiqiang Yang}
\email{d11102004@163.com}
\affiliation{Department of Physics, Liaoning Normal University, Dalian, 116029, People's Republic of China}

\author{David  F. Mota}
\email{mota@uio.no}
\affiliation{Institute of Theoretical Astrophysics, University of Oslo, P.O.Box 1029 Blindern, N-0315 Oslo, Norway}


\begin{abstract}

We investigate the dark energy phenomenology in an extended parameter space where we allow the curvature density of our universe as a free-to-vary parameter. The inclusion of the curvature density parameter is motivated from the recently released observational evidences indicating the closed universe model at many standard deviations.  Here we assume that the dark energy equation-of-state follows the PADE approximation, a generalized parametrization that may recover a variety of existing dark energy models.  Considering three distinct PADE parametrizations, labeled as PADE-I, SPADE-I and PADE-II, we first constrain the cosmological scenarios driven by them using the joint analyses of a series of recently available cosmological probes, namely, Pantheon sample of Supernovae Type Ia, baryon acoustic oscillations, big bang nucleosynthesis, Hubble parameter measurements from cosmic chronometers, cosmic microwave background distance priors from Planck 2018 and then we include the future Gravitational Waves standard sirens (GWSS) data from the Einstein telescope with the combined analyses of these current cosmological probes. We find that the current cosmological probes indicate a very strong evidence of a dynamical dark energy at more than 99\% CL in both  PADE-I, and PADE-II,  but no significant evidence for the non-flat universe is found in any of these parametrizations. Interestingly,  when the future GWSS data from the Einstein telescope are included with the standard cosmological probes an evidence of a non-flat universe is found in all three parametrizations together with a very strong preference of a dynamical dark energy at more than 99\% CL in both  PADE-I, and PADE-II. Although from the information criteria analysis, namely, AIC, BIC, DIC, the non-flat $\Lambda$-Cold Dark Matter model remains the best choice, however, in the light of DIC, PADE parametrizations are still appealing.

\end{abstract}

\maketitle

\section{Introduction}
\label{sec-introduction}

The physics of the dark sector of the universe has remained mysterious to the scientific community of modern cosmology.  The discovery of the late time cosmic acceleration has challenged the standard cosmology and expressed different views, such as the presence of some dark energy (DE) sector (hypothetical fluid having negative pressure in the context of Einstein's General Relativity), see  \cite{Sahni:1999gb,Peebles:2002gy,Copeland:2006wr,Sahni:2006pa,Frieman:2008sn,Bamba:2012cp} and the references therein, or the need of modification of the Einstein's General Relativity in various ways, see \cite{Nojiri:2006ri,Nojiri:2010wj,Capozziello:2011et,Clifton:2011jh,Koyama:2015vza,Cai:2015emx,Nojiri:2017ncd,CANTATA:2021ktz,Bahamonde:2021gfp} and the references therein. As a result, over the last couple of years, modern cosmology has witnessed the appearance of many theoretical models in the context of both DE and modified gravity theories, and these models have been confronted with a variety of astronomical probes \cite{Linder:2002et,Gorini:2002kf,Huterer:2002hy,Jimenez:2003iv,Zhang:2005hs,Capozziello:2005pa,Brookfield:2007au,Tsujikawa:2007xu,Li:2009mf,Brax:2009ab,Li:2010re,Li:2010zw,Li:2011wu,Martinelli:2011wi,Paliathanasis:2014zxa,DiValentino:2015ola,Sharov:2015ifa,vandeBruck:2016hpz,Odintsov:2017qif,Mifsud:2017fsy,Yang:2017alx,Zhao:2017cud,VanDeBruck:2017mua,Yang:2018qec,Odintsov:2018qug,Yang:2018pej,Arnold:2019vpg,Yang:2019jwn,Cai:2019bdh,Li:2019san,Martinelli:2019dau,Anagnostopoulos:2019miu,DiValentino:2019jae,Li:2020ybr,Pan:2020bur,Odintsov:2020voa,Anagnostopoulos:2021ydo,Guo:2021rrz,Gariazzo:2021qtg,Sharov:2022hxi,Wang:2021kxc,Staicova:2021ntm,Saridakis:2021xqy,Nunes:2022bhn,Odintsov:2022zrj,Kumar:2022nvf,Tsedrik:2022cri,Zhai:2023yny,Betts:2023hqo}. However, the intrinsic mechanism behind this late-time cosmic acceleration is still unknown and the scientific community is still wondering whether DE or modified gravity or some other alternative theory is the ultimate answer to this observed phenomenon. In addition, the emergence of tensions in the key cosmological parameters \cite{Abdalla:2022yfr,DiValentino:2021izs,Perivolaropoulos:2021jda,Schoneberg:2021qvd} and the possibility of a non-flat universe (specifically the preference of a closed universe model) indicated by the recent observational evidences  \cite{Aghanim:2018eyx,Handley:2019tkm,DiValentino:2019qzk,DiValentino:2020hov} demand that the cosmological models deserve further attention.

In the present article we focus on the DE phenomenology with a special attention to the curvature of our universe. 
The spatial curvature of the universe is commonly assumed to be zero in the analyses of the cosmological models. This is an assumption and such assumption is motivated based on the facts that the dynamics of  a  cosmological model becomes simple for a zero spatial curvature of the universe, and secondly the astronomical data in the past reported a very small value of the curvature density parameter  \cite{Gaztanaga:2008de,Mortonson:2009nw,Suyu:2013kha,LHuillier:2016mtc}.  However, recent observational data have expressed a different view on the spatial curvature of the universe~\cite{Aghanim:2018eyx,Handley:2019tkm,DiValentino:2019qzk,DiValentino:2020hov} and the debate on the curvature has increased in the community.  According to the evidences from the data, the analyses with the base {\it Plik}
likelihood of the cosmic microwave background (CMB)  temperature and polarization spectra from Planck 2018 \cite{Aghanim:2018eyx,DiValentino:2019qzk,Handley:2019tkm,DiValentino:2020hov} intimate that our universe is closed at more than 99\% CL. Particularly, in the context of a non-flat $\Lambda$CDM model, the present value of the curvature density parameter is found to be $\Omega_{k0} =-0.044^{+0.033}_{-0.034}$ at 95\% CL by Planck TT, TE, EE+lowE \cite{Aghanim:2018eyx}. The inclusion of lensing with Planck TT, TE, EE+lowE although reduces the evidence but still $\Omega_{k0}$ remains non-zero at more than 95\% CL ($\Omega_{k0} = -0.011^{+0.013}_{-0.012}$ at 95\% CL for Planck TT, TE, EE+lowE+lensing \cite{Aghanim:2018eyx}). However, for  Planck TT, TE, EE+lowE+lensing+BAO, the evidence for the curvature becomes weak and $\Omega_{k0}$ becomes consistent with zero ($\Omega_{k0} = 0.0007 \pm 0.0037$ at 95\% CL) \cite{Aghanim:2018eyx}. On the other hand, using the alternative \emph{CamSpec} likelihood of CMB from Planck 2018 and considering the $\Lambda$CDM model in the background with the free-to-vary curvature density parameter, the evidence for the curvature is slightly reduced for Planck TT, TE, EE+lowE ($\Omega_{k0} =-0.037^{+0.032}_{-0.034}$ at 95\% CL) \cite{Aghanim:2018eyx}, but $\Omega_{k0}$ remains non-zero at more than 95\% CL. However, similar to the {\it Plik} likelihood case, in this case too, the inclusion of lensing and baryon acoustic oscillations with Planck TT, TE, EE+lowE makes the evidence of the curvature weak and the flat universe is preferred: $\Omega_{k0} = -0.011^{+0.012}_{-0.013}$ at 95\% CL for Planck TT, TE, EE+lowE+lensing \cite{Aghanim:2018eyx} and $\Omega_{k0} = 0.0005^{+0.0038}_{-0.0040}$ at 95\% CL for Planck TT, TE, EE+lowE+lensing+BAO \cite{Aghanim:2018eyx}. We further refer to Refs.~\cite{Efstathiou:2019mdh,Efstathiou:2020wem} where the authors performed a detailed investigations on the universe's curvature using the \emph{CamSpec} likelihood of CMB from Planck 2018. In particular, the authors of \cite{Efstathiou:2019mdh} created different \emph{CamSpec} likelihoods, and found that for the 12.1HM likelihood, Planck TT, Planck TE, Planck TTTEEE, Planck TTTEEE+lensing, indicate a preference of a closed universe \cite{Efstathiou:2019mdh}, while for the 12.5MHcl likelihood, except for Planck TE, the results from Planck TT, Planck TTTEEE, Planck TTTEEE+lensing also support the closed universe scenario~\cite{Efstathiou:2019mdh}. However, as further noticed in \cite{Efstathiou:2019mdh,Efstathiou:2020wem}, the evidence for non-zero curvature of the universe goes away when  some external probes, such as baryon acoustic oscillations are added to CMB from Planck. Thus, from the observational reports, it is difficult to remove the curvature of the universe 
from the picture.   
It has been observed that the curvature term if  plugged into the gravitational equations, then it could significantly affect the cosmological parameters \cite{Dossett:2012kd}.  So, the small values of the curvature density parameter obtained from different observational data \cite{Gaztanaga:2008de,Mortonson:2009nw,Suyu:2013kha,LHuillier:2016mtc}, do not justify the assumption of the spatial flatness of the universe as there is no strong theoretical background available in this regard~\cite{Anselmi:2022uvj}. In fact, it is very natural to consider the curvature of the universe into the cosmic picture for completeness and then investigate its consequences on the dynamics of the universe. Thus, based on the facts that, the non-zero spatial curvature of the universe is indicated by recent observations \cite{Aghanim:2018eyx,Handley:2019tkm,DiValentino:2019qzk,DiValentino:2020hov,Efstathiou:2020wem}; no solid theoretical background is available to avoid the curvature of the universe  \cite{Anselmi:2022uvj}; the presence of a spatial curvature of the universe in the gravitational equations could affect the cosmological parameters \cite{Dossett:2012kd};   one can infer that a thorough cosmological description might be obtained by allowing the spatial curvature in the gravitational equations and test the underlying cosmological scenario with the available data even if the resulting picture becomes complicated. According to the past  and current records in the literature, the curvature of our universe has remained one of the hot topics in cosmology
\cite{Virey:2008nu,Li:2012vn,Farooq:2016zwm,Park:2017xbl,DiValentino:2020kpf,DiValentino:2020hov,Benisty:2020otr,Vagnozzi:2020rcz,Vagnozzi:2020dfn,Yang:2021hxg,Cao:2021ldv,Gonzalez:2021ojp,Dhawan:2021mel,Fondi:2022tfp,DiValentino:2022oon,DiValentino:2022rdg,Yang:2022kho,Stevens:2022evv,Glanville:2022xes,Bel:2022iuf,Cao:2023eja,Favale:2023lnp}. Therefore, without going into the ongoing debate $-$ `whether our universe is spatially flat or curved', in this article we re-examine some DE models in presence of the spatial curvature of the universe and allow the observational data to decide its fate.

 Now focusing on the DE sector, even though the recent observational evidences are in favor of the cosmological constant \cite{Escamilla:2023oce}, but a mild inclination of DE towards the phantom regime needs further attention as indicated by the authors of \cite{Escamilla:2023oce}. On the other hand, according to the model independent reconstruction of the DE equation of state \cite{Zhao:2017cud,Zhang:2019jsu} the possibility of a dynamical DE featuring a very a mild deviation from the cosmological constant is very hard to exclude. 
The dynamical DE is one of the simplest extensions of the $\Lambda$CDM cosmology (see \cite{Chevallier:2000qy,Linder:2002et,Upadhye:2004hh,Gong:2009ye,Casarini:2010uc,Tsujikawa:2012hv,Novosyadlyj:2013nya,Akarsu:2015yea,DiValentino:2016hlg,DiValentino:2017zyq,DiValentino:2017rcr,Yang:2018qmz,DiValentino:2019dzu,Vagnozzi:2018jhn,Du:2018tia,Yang:2018prh,Park:2018fxx,Benevento:2020fev,DiValentino:2020naf,Alestas:2020zol,Yang:2021flj,Li:2019yem,Pan:2019hac} and the references therein), and it has been observed that the dynamical DE models could 
play a very crucial role in easing the Hubble constant tension \cite{DiValentino:2016hlg,DiValentino:2017zyq,DiValentino:2017rcr,Yang:2018qmz,DiValentino:2019dzu,Benevento:2020fev,DiValentino:2020naf,Alestas:2020zol,Yang:2021flj,
Li:2019yem,Pan:2019hac}. 
However, there is no fundamental guiding principle available in the literature yet to determine the correct equation of state of DE. In this article we adopt a very general avenue in which the equation of state of DE can be constructed from the PADE parametrization~\cite{ASENS_1892_3_9__S3_0}.  
The PADE parametrization has received notable interest  in the cosmology community \cite{Gruber:2013wua,Wei:2013jya,Aviles:2014rma,Liu:2014vda,Zhou:2016nik,Rezaei:2017yyj,Capozziello:2017ddd,Mehrabi:2018oke,Rezaei:2019hvb}. This particular parametrization has a very rich character in the sense that it allows one to construct a cluster of DE parametrizations. The well known Taylor series expansion is a special case of the 
PADE parametrization and thus the DE parametrizations originating from the Taylor expansion are special cases of the PADE parametrization. According to the past historical records, the PADE parametrizations have been investigated in a curvature free universe. However, given the recent interests on the curvature of the universe \cite{Aghanim:2018eyx,Handley:2019tkm,DiValentino:2019qzk,DiValentino:2020hov,Efstathiou:2020wem} (also see \cite{Virey:2008nu,Li:2012vn,Farooq:2016zwm,Park:2017xbl,DiValentino:2020kpf,DiValentino:2020hov,Benisty:2020otr,Vagnozzi:2020rcz,Vagnozzi:2020dfn,Yang:2021hxg,Cao:2021ldv,Gonzalez:2021ojp,Dhawan:2021mel,Fondi:2022tfp,DiValentino:2022oon,DiValentino:2022rdg,Yang:2022kho,Stevens:2022evv,Glanville:2022xes,Bel:2022iuf,Cao:2023eja,Favale:2023lnp}), in this work we take an attempt to investigate some recently explored DE models originating from the PADE parametrization \cite{Rezaei:2017yyj}. We constrain the models  using the combined probes from the Pantheon sample of Supernovae Type Ia, Baryon acoustic oscillations distance measurements, information from big bang nucleosynthesis, Hubble parameter measurements from the cosmic chronometers and the CMB distance priors from Planck 2018 final release. Our analyses reveal that the DE models constructed from the PADE parametrization deserve further attention in the cosmology community.

The article is structured as follows. In \autoref{sec-basic-eqns} we describe the key gravitational equations, the PADE parametrization 
and the DE parametrizations constructed from the PADE parametrization that we wish to study. In \autoref{sec-data} we discuss the observational datasets that have been used to constrain the proposed DE parametrizations. Then in \autoref{sec-results} we discuss the results extracted out of these DE parametrizations. Finally, in \autoref{sec-conclusion} we close this article with its key results.

\section{Dynamics of DE: PADE parametrization}
\label{sec-basic-eqns}

We consider a homogeneous and isotropic universe characterized by the Friedmann-Lema\^{i}tre-Robertson-Walker  as follows
line element given by  
 \begin{eqnarray}\label{flrw}
 ds^2 = -dt^2 + a^2 (t) \left[\frac{dr^2}{1-kr^2} + r^2 (d\theta^2 + \sin^2 \theta d \phi^2)\right]
 \end{eqnarray}
 where $(t, r, \theta, \phi)$ are the co-moving coordinates, $a(t)$
 is the expansion scale factor of the universe and $k$ denotes the spatial geometry of the universe which for $k =0$, $+1$, $-1$, respectively denotes a spatially flat, closed and an open universe.
 We further assume that the matter sector of the universe comprising  radiation, pressure-less matter (baryons plus cold dark matter) and DE, is minimally coupled to gravity which follows Einstein's GR and the fluid components comprising the entire matter sector do not interact with one another, that means, no energy exchange between any two fluids is allowed. Having this description above, one can write down the Einstein's gravitational equations in the FLRW background (\ref{flrw}) as follows 

\begin{eqnarray}
H^2 + \frac{k}{a^2}=\frac{8\pi G}{3}(\rho_{\rm b}+ \rho_{\rm rad}+\rho_{\rm dm}+\rho_{\rm de})\;\label{efe1},\\
2 \dot{H} + 3 H^2 + \frac{k}{a^2} = - 8 \pi G (p_{\rm b}+ p_{\rm rad}+p_{\rm dm}+p_{\rm de}),\label{efe2}
\end{eqnarray}
where $\rho_i$, $p_i$ ($i =  {\rm b}, {\rm rad}, {\rm dm}, {\rm de}$) are respectively the energy density and pressure of the $i$-th fluid where  ${\rm b}, {\rm rad}, {\rm dm}, {\rm de}$ respectively denotes the baryons, radiation, pressure-less DM and the DE sector. In terms of the critical density $\rho_c = 3H^2/8 \pi G$, the Hubble equation  (\ref{efe1}) can be recast into $1 = \Omega_k + \sum_i \Omega_i$, where $\Omega_i = \rho_i/\rho_c$ denotes the density parameter of the $i$-th fluid  and $\Omega_k = -k/(a^2 H^2)$ is  the curvature density parameter\footnote{Thus, one may note that for $k =0$ (flat universe), $\Omega_k =0$,  for $k =1$ (closed universe) $\Omega_k < 0$, and for $k = -1$, $\Omega_k > 0$. }. As the fluid components do not interact with each other,  therefore, the conservation equations for radiation, matter and DE can be expressed as

\begin{eqnarray}\label{cons-eqn}
&&\dot{\rho}_{\rm b} + 3H\rho_{\rm b}=0\;,\label{baryons}\\
&& \dot{\rho}_{\rm rad} + 4H\rho_{\rm rad}=0\;,\label{radiation}\\
&&\dot{\rho}_{\rm dm} + 3H\rho_{\rm dm}=0\;,\label{matter}\\
&&\dot{\rho}_{\rm de} + 3H(1+w_{\rm de})\rho_{\rm de}=0\;\label{de},
 \end{eqnarray}
where $w_{\rm b} =0$, $w_{\rm rad} = 1/3$, $w_{\rm dm} =0$ (pressure-less matter) and $w_{\rm de}$ are the equation-of-state (EoS) parameters representing the baryons, radiation, matter and the DE sector. 
From the conservation equations, one can see that, $\rho_{\rm b} \propto (a/a_0)^{-3}$, $\rho_{\rm rad} \propto (a/a_0)^{-4}$, $\rho_{\rm dm} \propto (a/a_0)^{-3}$ (here $a_0$ refers to the current value of the scale factor) and the DE sector evolves as 

\begin{eqnarray}
\rho_{\rm de} = \rho_{\rm de,0} \exp \left[ 3 \; \int_{a}^{a_0} \frac{1+w_{\rm de}(a^\prime)}{a'} da^\prime \right],
\end{eqnarray}
 where $\rho_{\rm de,0}$ denotes the present value of the DE density. Thus, the dynamics of DE entirely depends on its EoS and we note that there is no unique choice of the EoS of DE. Here we focus on a generalized EoS of DE that follows the PADE parametrization. The PADE approximation  of any arbitrary  function $f$ of order $(m, n)$ reads \cite{ASENS_1892_3_9__S3_0} (also see \cite{baker_graves-morris_1996})

\begin{eqnarray}\label{PADE}
f(x)=\frac{c_0+c_1x+c_2x^2+...+c_m x^m}{d_0+d_1x+d_2x^2+...+d_n x^n},
\end{eqnarray}
where the exponents $(m, n)$ are positive and the coefficients $c_{\rm i}$'s ($i = 1, 2,...m$), $d{\rm _j}$'s ($j = 1, 2, 3,...n$) are any real numbers.  The PADE approximation (\ref{PADE}) is a very phenomenologically rich parametrization which for $d_{\rm j}=0$ where $j \geq 1$, recovers
the standard Taylor expansion of $f(x)$. In the cosmological context, we assume that DE EoS follows  the PADE approximation 
as follows 
\begin{eqnarray}\label{PADE-w}
w_{de} (x)=\frac{c_0+c_1x+c_2x^2+...+c_m x^m}{d_0+d_1x+d_2x^2+...+d_n x^n},
\end{eqnarray}
where $x$ could be any  cosmological variable,  for instance, the scale factor $a$ of the FLRW universe, the cosmic time $t$ or maybe the Hubble rate $H$ of the FLRW universe. Now, for a variety of $(m, n)$,  as well as the parameters $c_{\rm i}$, $d_{\rm j}$, one can generate a cluster of $w_{de}$ parametrizations. One can quickly identify that the well known Chevallier-Polarski-Linder parametrization \cite{Chevallier:2000qy,Linder:2002et} is a special case of the PADE approximation. In fact, there are many EoS of DE in the literature that can be directly recovered from this general parametrization.

\subsection{Parametrization I}

The first parametrization in this series goes as follows. Considering $x = 1-a/a_0$, we take the expansion of $w_{de} (x)$ up to order $(1, 1)$

\begin{equation}\label{pade1}
w_{\rm de}(a)=\frac{c_0+c_{1}(1-a/a_0)}{d_0+d_{1}(1-a/a_0)}\;.
\end{equation}
If $d_0 \neq 0$, then eqn. (\ref{pade1}) can be expressed as 
\begin{equation}\label{pade1.1}
w_{\rm de}(a)=\frac{w_0 +w_{1}(1-a/a_0)}{1+w_{2}(1-a/a_0)}\;.
\end{equation}
where $w_0 = c_0/d_0$, $w_1 = c_1/d_0$ and $w_2 =d_1/d_0$. We label this parametrization as ``PADE-I''. One can quickly identify  that $w_0$ is actually the current value of the DE equation of state.

\subsection{Parametrization II}

The second parametrization of this series is obtained by setting $w_1 = 0$ in PADE-I of eqn. (\ref{pade1.1}) and this reads 

\begin{equation}\label{simplified-pade1.1}
w_{\rm de}(a)=\frac{w_0}{1+w_{2}(1-a/a_0)}\;.
\end{equation}
which we call as the Simplified PADE-I (labeled as ``SPADE-I'') as in Ref. \cite{Rezaei:2017yyj}. This parametrization has one less parameter compared to ``PADE-I'' parametrization.

\subsection{Parametrization III}

For the last parametrization, we choose $x$ differently. Instead of 
eqn. (\ref{pade1.1}), where we took the PADE parametrization of (\ref{PADE}) in terms of $ x = (1-a/a_0)$,  
we consider $x = \ln (a/a_0) $ in (\ref{PADE}) up to $(m, n) = (1, 1)$ 
and get the following parametrization 

\begin{equation}\label{pade2}
w_{\rm de}(a)=\frac{w_0+w_{1}\ln{(a/a_0)}}{1+w_{2}\ln{(a/a_0)}}\;,
\end{equation}
which we label as ``PADE-II''.
Considering these three parametrizations representing the DE EoS, we 
fitted the cosmological scenarios using various observational datasets. Having these parametrizations for DE in terms of its EoS, in the following we describe the observational datasets that have been used in this article. 

\section{Observational data}
\label{sec-data}

In this section we describe the observational datasets used  to constrain the underlying cosmological scenarios.

\subsection{Current cosmological probes}
The current cosmological probes that we used in this work are as follows: 

\begin{itemize}

\item {\bf Supernovae Type Ia:} We have used the Pantheon sample of Supernovae Type Ia which involves the 1048 data points for the apparent magnitude of SNIa within the redshift range $0.01 < z < 2.26$ \cite{Scolnic:2017caz}. We label this dataset as ``SN''.

\item {\bf Baryon acoustic oscillations:} Concerning  the baryonic acoustic oscillations (BAOs) data we have used the radial component of the anisotropic BAOs extracted from the measurements of the power spectrum and bispectrum from the BOSS, Data Release 12, galaxy sample \cite{Gil-Marin:2016wya}, the complete SDSS III Ly$\alpha$-quasar \cite{duMasdesBourboux:2017mrl} and the SDSS-IV extended BOSS DR14 quasar sample from \cite{Gil-Marin:2018cgo}. We label this dataset as ``BAO''.

\item {\bf Big Bang Nucleosynthesis:} We have used the baryon density information from the Big Bang Nucleosynthesis where $\Omega_bh^2 = 0.022 \pm 0.002$ (at 95\% CL) \cite{Burles:2000zk}. We label this data point as ``BBN''.

\item {\bf  Hubble parameter measurements from cosmic chronometers:} The cosmic chronometers data, namely the data points on the Hubble rate, $H(z)$, obtained from the spectroscopic techniques applied to passively–evolving galaxies, i.e., galaxies with old stellar populations and low star formation rates. We have used 37 data points of this type in the present work from Refs. \cite{Jimenez:2003iv,Simon:2004tf,Stern:2009ep,Zhang:2012mp,Moresco:2012jh,Moresco:2015cya,Moresco:2016mzx,Ratsimbazafy:2017vga}. We label this dataset as ``$H (z)$''.

\item {\bf Cosmic microwave background:} We have used the cosmic microwave background (CMB) distance priors from \cite{Chen:2018dbv}. In this reference, authors obtained the distance priors in several cosmological models using Planck 2018 TT,TE,EE + lowE, which is the latest CMB data from the final data release of the Planck team \citep{Planck:2018vyg}. In particular, we use the distance priors $(l_A,~R,~\Omega_bh^2)$ for the flat $\Lambda$CDM model  (see Table I of \cite{Chen:2018dbv}) in which $l_A$ is the angular scale of the sound horizon at recombination which determines the acoustic peak structure of the CMB,  $R$ is the scaled distance to recombination which determines the amplitude of the acoustic peaks in the power spectrum of CMB temperature anisotropy and $\Omega_b h^2$ is the physical baryon density. We label this data as ``CMB''. 

\end{itemize}

\subsection{GWSS luminosity distance measurements}

The mock GWSS luminosity distance measurements used in this work have been generated by matching the expected sensitivity of the Einstein Telescope, a proposed ground based third-generation GW detector~\cite{Sathyaprakash:2012jk,Maggiore:2019uih}. After 10 years of full operation, Einstein Telescope is likely to detect $\mathcal{O} (10^3)$ GWSS events, while as argued in \cite{Maggiore:2019uih}, the number of detections might be low in reality. However, following the earlier works in this direction~\cite{Zhao:2010sz,Cai:2016sby,Wang:2018lun,Du:2018tia,Yang:2019vni,Yang:2019bpr,Yang:2020wby,Matos:2021qne,Pan:2021tpk}, we work with the  $\mathcal{O} (10^3)$ detections by the Einstein Telescope.  The methodology to generate the mock GWSS dataset can be found in detail in Refs. \cite{Du:2018tia,Yang:2019bpr,Yang:2019vni}. Here we briefly describe the key points that are essential for this article. 
The first step to generate this mock GWSS dataset relies on the identification of the GW sources.  We consider that a GW event is detected from the following two distinct binary systems: a combination of a Black Hole (BH) and a Neutron Star (NS) merger, labeled as BHNS and the binary neutron star (BNS) merger. Then we determine the merger rate $R(z)$ of the sources and using the merger rate of the sources, the  redshift distribution of the  sources, $P (z)$  can be found to be~\cite{Sathyaprakash:2009xt,Zhao:2010sz,Cai:2016sby,Wang:2018lun,Du:2018tia,Yang:2019bpr}

\begin{eqnarray}
    P (z) \propto \frac{4 \pi d_C^2 (z) R (z)}{H (z) (1+z)},
\end{eqnarray}
where $d_C (z) \equiv \int_{0}^{z} H^{-1} (z^\prime) dz^\prime$ denotes the co-moving distance, and for the merger rate, following piece-wise linear function estimated in~\cite{Schneider:2000sg} (see also Refs~\cite{Cutler:2009qv,Zhao:2010sz,Cai:2016sby,Wang:2018lun,Du:2018tia,Yang:2019bpr}) has been considered:  $R(z) = 1+2 z$ for $z\leq 1$, $R (z) = \frac{3}{4}(5-z)$,  for $1<z<5$ and $R (z) = 0$ for $z > 5$.  Having the merger rate and henceforth $P (z)$, we sample 1000
triples ($z_i$, $d_{L} (z_i)$, $\sigma_i$) where $z_i$ is the redshift of a GW source, $d_L (z_i)$ denotes the measured luminosity distance at redshift $z_i$ and $\sigma_i$ refers to the uncertainty of the luminosity distance $d_L (z_i)$. 

As the next step we choose a fiducial model because the redshift distribution of the sources, $P (z)$, includes  the expansion history $H(z)$ corresponding to the fiducial model. In this work we choose the flat $\Lambda$CDM as the fiducial model\footnote{Although in this work we have assumed the flat $\Lambda$CDM model as the fiducial model, but it will be interesting to investigate the impact of the mock GWSS dataset under the assumption of the non-flat $\Lambda$CDM cosmology as the fiducial model. However, since this is a mock dataset, therefore, we have considered a very minimal cosmological scenario as the fiducial one which is the flat $\Lambda$CDM. } and took the estimated best-fit values of the model parameters from Planck 2018 \cite{Planck:2018vyg}. 
Now, for the fiducial model, the luminosity distance at the redshift $z_i$ can be estimated as 

\begin{eqnarray}
d_L(z_i) = (1+z_i)\int_0^{z_i}\frac{dz'}{H(z')}\,.
\label{eq:luminosity}
\end{eqnarray}
Finally, we determine the uncertainty associated with this luminosity distance measurements (for technical details on the error estimation, we refer to Refs. \cite{Zhao:2010sz,Cai:2016sby,Wang:2018lun,Du:2018tia,Yang:2019bpr}). 
The luminosity distance measurement $d_L (z_i)$ has two kind of uncertainties $-$ the instrumental uncertainty $\sigma_i^{\rm inst}$  and the weak lensing uncertainty $\sigma_i^{\rm lens}$. The instrumental error can be estimated as $\sigma_i^{\rm inst} ~(\simeq 2 d_L (z_i)/\mathcal{S}$ where $\mathcal{S}$ refers to the combined signal-to-noise ratio of the Einstein Telescope) using the Fisher matrix approach and assuming that $d_L (z_i)$ is uncorrelated with the uncertainties on the remaining GW parameters   (see~\cite{Zhao:2010sz,Cai:2016sby,Wang:2018lun,Du:2018tia,Yang:2019bpr}). The lensing error is  $\sigma_i^{\rm lens} \simeq 0.05 z_i d_L (z_i)$~\cite{Zhao:2010sz}. The total uncertainty on $d_L (z_i)$ is determined as  $\sigma_i = \sqrt{(\sigma_i^{\rm inst})^2 + (\sigma_i^{\rm lens})^2}$.  Last but not least, we make a note that the combined signal-to-noise ratio of the GW detector, i.e. $\mathcal{S}$ is an important quantity because for the Einstein Telescope, $\mathcal{S}$ should be at least $8$ for a GW detection~\cite{Sathyaprakash:2009xt}.

In our analysis, for each of the parameterizations, we have the following free parameters:

\begin{itemize}

    \item PADE$-$I: $\Omega_{\rm b0}$, $\Omega_{\rm dm0}$, $\Omega_{k0}$, $h$, $w_0$, $w_1$, $w_2$

     \item SPADE$-$I: $\Omega_{\rm b0}$, $\Omega_{\rm dm0}$, $\Omega_{k0}$, $h$, $w_0$, $w_2$

      \item PADE$-$II: $\Omega_{\rm b0}$, $\Omega_{\rm dm0}$, $\Omega_{k0}$, $h$, $w_0$, $w_1$, $w_2$

       \item $\Lambda$CDM: $\Omega_{\rm b0}$, $\Omega_{\rm dm0}$, $\Omega_{k0}$, $h$
       
\end{itemize}
where note that $h = H_0/100$. Having all the datasets presented above, we perform  the likelihood analysis using the Markov Chain Mote Carlo (MCMC) method aiming to constrain all the parameters of the proposed cosmological scenarios. Adopting the Metropolis algorithm, we let each of the model parameters to vary in the parameter space with no limitation (i.e. each of the parameter is varied in a wide region). 
In each of the model analysis, we have performed 200000 chains to obtain the best-fit values of the free parameters. Using the unlimited parameter space and large number of chains in MCMC, we can avoid the risk of picking out local best fit values in the parameters space. Following the approaches described in Refs. \cite{Rezaei:2019roe,Rezaei:2020lfy,Rezaei:2021qwd}, we have examined different initial values and priors on each of the free parameters. We observed that our results are independent on the selected initial values and the prior of the free parameters. In what follows we present and discuss the main results of this article.  

\begin{table*}
\centering
\caption{Summary of the 68\% CL constraints on various free parameters of the cosmological scenarios in the non-flat universe, namely, PADE-I, SPADE-I, PADE-II and $\Lambda$CDM for  the combined datasets SN+BAO+BBN+$H(z)$ and SN+BAO+BBN+$H(z)$+CMB. Note that $\Omega_{m0}$ denotes the total density parameter for the matter sector at present time, that means, $\Omega_{\rm m0} = \Omega_{\rm b0}+\Omega_{\rm dm0}$. }
 \begin{tabular}{c  c  c c c c c }
 \hline 
 Model & $\Omega_{m0}$  & $\Omega_{k0}$  & $h$ [km/s/Mpc]  & $w_0$   & $w_1$ 
 & $w_2$\\
 \hline 

\multicolumn{7}{c}{SN+BAO+BBN+$H(z)$} \\
 \hline 

 PADE-I & $0.2682\pm 0.0090$ & $0.004^{+0.026}_{-0.031}$ & $0.6868\pm 0.0082$  & $-0.926\pm 0.057$ & $-0.551\pm 0.086$  & $1.90\pm 0.15$  \\
 
  SPADE-I & $0.262\pm 0.016$ & $0.05^{+0.11}_{-0.13}$  & $0.685\pm 0.010$ & $-1.018\pm 0.099$ & $-$ & $0.67\pm 0.59$\\

 PADE-II & $0.254\pm 0.013$ & $ -0.041\pm 0.053$ & $0.6812^{+0.0088}_{-0.0098}$ & $-1.085\pm 0.062$ & $1.60^{+0.16}_{-0.22}$  & $-2.01^{+0.26}_{-0.20}$ \\
 
  $\Lambda$CDM & $0.271\pm 0.015$ & $-0.017\pm 0.038$ & $0.6869\pm 0.0097$ & $-$ &  $-$ & $-$ \\

\hline 
\multicolumn{7}{c}{SN+BAO+BBN+$H(z)$+CMB} \\
 \hline 
PADE-I & $0.2680\pm 0.0079$ & $0.011^{+0.023}_{-0.028}$ & $0.6974\pm 0.0074$  & $-1.036\pm 0.059$ & $-0.480\pm 0.029$  & $1.81\pm 0.10$  \\
 
 SPADE-I & $0.2765\pm 0.0088$ & $-0.006\pm 0.037$  & $0.6942\pm 0.0080$ & $-1.006^{+0.058}_{-0.051}$ & $-$ & $0.06^{+0.31}_{-0.28}$ \\
 
 PADE-II & $0.2762\pm 0.0090$ & $0.017^{+0.043}_{-0.051}$ & $0.6951\pm 0.0082$ & $-1.040\pm 0.067$ & $1.89\pm 0.29$  & $-2.35\pm 0.33$ \\
 
 $\Lambda$CDM & $0.2776\pm 0.0087$ & $-0.008\pm 0.022$ & $0.6944\pm 0.0080$ & $-$ &  $-$ & $-$ \\
\hline 
\hline 
\end{tabular}\label{tab:best}
\end{table*}
\begin{table*}
\centering
\caption{Summary of the 68\% CL constraints on various free parameters of the cosmological scenarios in flat space, namely, PADE-I, SPADE-I, PADE-II and $\Lambda$CDM for the combined datasets SN+BAO+BBN+$H(z)$ and SN+BAO+BBN+$H(z)$+CMB. Note that $\Omega_{m0}$ denotes the total density parameter for the matter sector at present time, that means, $\Omega_{\rm m0} = \Omega_{\rm b0}+\Omega_{\rm dm0}$.}
 \begin{tabular}{c  c  c  c c c }
 \hline 
 Model & $\Omega_{m0}$  & $h$ [km/s/Mpc]  & $w_0$   & $w_1$ 
 & $w_2$\\
 \hline 

\multicolumn{6}{c}{SN+BAO+BBN+$H(z)$} \\
 \hline 

 PADE-I & $0.254\pm 0.009$  & $0.683\pm 0.0080$  & $-1.028\pm 0.062$ & $0.582\pm 0.093$  & $-1.03\pm 0.092$  \\
 
 SPADE-I & $0.263\pm 0.018$ & $0.685\pm 0.011$ & $-1.067\pm 0.088$ & $-$ & $0.676\pm 0.62$\\

 PADE-II & $0.259\pm 0.015$ &  $0.689\pm 0.010$ & $-1.027\pm 0.058$ & $0.0639^{+0.09}_{-0.17}$  & $-0.464^{+0.21}_{-0.20}$ \\
 
 $\Lambda$CDM & $0.276\pm 0.014$ & $0.689\pm 0.0094$ & $-$ &  $-$ & $-$ \\

\hline 
\multicolumn{6}{c}{SN+BAO+BBN+$H(z)$+CMB} \\
 \hline 
PADE-I & $0.286\pm 0.0074$  & $0.693\pm 0.0071$  & $-0.932\pm 0.056$ & $-0.219\pm 0.019$  & $-0.179\pm 0.10$  \\
 
 SPADE-I & $0.273\pm 0.0084$ & $0.693\pm 0.0078$ & $-1.054\pm 0.051$ & $-$ & $0.44^{+0.03}_{-0.02}$ \\
 
 PADE-II & $0.279\pm 0.0086$  & $0.691\pm 0.0080$ & $-0.994\pm 0.068$ & $0.167\pm 0.021$  & $-0.096\pm 0.021$ \\
 
 $\Lambda$CDM & $0.277\pm 0.0084$ & $0.694\pm 0.0082$ & $-$ &  $-$ & $-$ \\
\hline 
\hline 
\end{tabular}\label{tab:flat1}
\end{table*}
\begin{table*}
	\centering
	\caption{The table shows how the inclusion of the recent SH0ES prior ($H_0 = 73.04 \pm 1.04 $ km/s/Mpc at 68\% CL) \cite{Riess:2021jrx} to both the combined datasets SN+BAO+BBN+$H(z)$ and SN+BAO+BBN+$H(z)$+CMB may affect the reduced Hubble constant $h$ obtained in the present cosmological scenarios. }
 \begin{tabular}{c c c c c c}
 \hline 
\multicolumn{6}{c}{Non-flat case}\\
 \hline
& \multicolumn{2}{c}{SN+BAO+BBN+$H(z)$} & & \multicolumn{2}{c}{SN+BAO+BBN+$H(z)$+CMB}\\
 \hline 
  & without $H_0$  &  with $H_0$  &   & without $H_0$  &  with $H_0$  \\
 \hline
 PADE-I & $0.6868 \pm 0.0082$  & $0.7074\pm 0.0076$  &   & $0.6974 \pm 0.0074$ & $0.7024\pm 0.0071$ \\
SPADE-I & $0.6850\pm 0.0100$  & $0.7100\pm 0.0077$  &   & $ 0.6942 \pm 0.0080$ & $0.7042\pm 0.0066$ \\
PADE-II & $0.6812^{+0.0088}_{-0.0098}$  & $0.7050\pm 0.0073$  &   & $0.6951\pm 0.0082$ & $0.7088\pm 0.0072$ \\
 $\Lambda$CDM & $0.6869\pm0.0097$ & $0.7048\pm 0.0073$ &  & $0.6944\pm 0.0080$ & $0.7070\pm 0.0075$ \\
 
 \hline
\multicolumn{6}{c}{Flat case}\\
 \hline
& \multicolumn{2}{c}{SN+BAO+BBN+$H(z)$} & & \multicolumn{2}{c}{SN+BAO+BBN+$H(z)$+CMB}\\
 \hline 
  & without $H_0$  &  with $H_0$  &   & without $H_0$  &  with $H_0$  \\
 \hline
 PADE-I & $0.6831 \pm 0.0080$  & $0.7029\pm 0.0078$  &   & $0.6932\pm 0.0071$ & $0.7062\pm 0.0071$ \\
SPADE-I & $0.6852\pm 0.0110$  & $0.7058\pm 0.0077$  &   & $0.6934\pm 0.0078$ & $0.7066\pm 0.0066$ \\
PADE-II & $0.6890\pm 0.0102$  & $0.7061\pm 0.0073$  &   & $0.6917\pm 0.0080$ & $0.7045\pm 0.0072$ \\
 $\Lambda$CDM & $0.6894\pm0.0094$ & $0.7059\pm 0.0073$ &  & $0.6942\pm 0.0082$ & $0.7063\pm 0.0075$ \\
 
 \hline
\end{tabular}
\label{table-H0}
\end{table*}
\begin{figure*}
    \includegraphics[width=10cm]{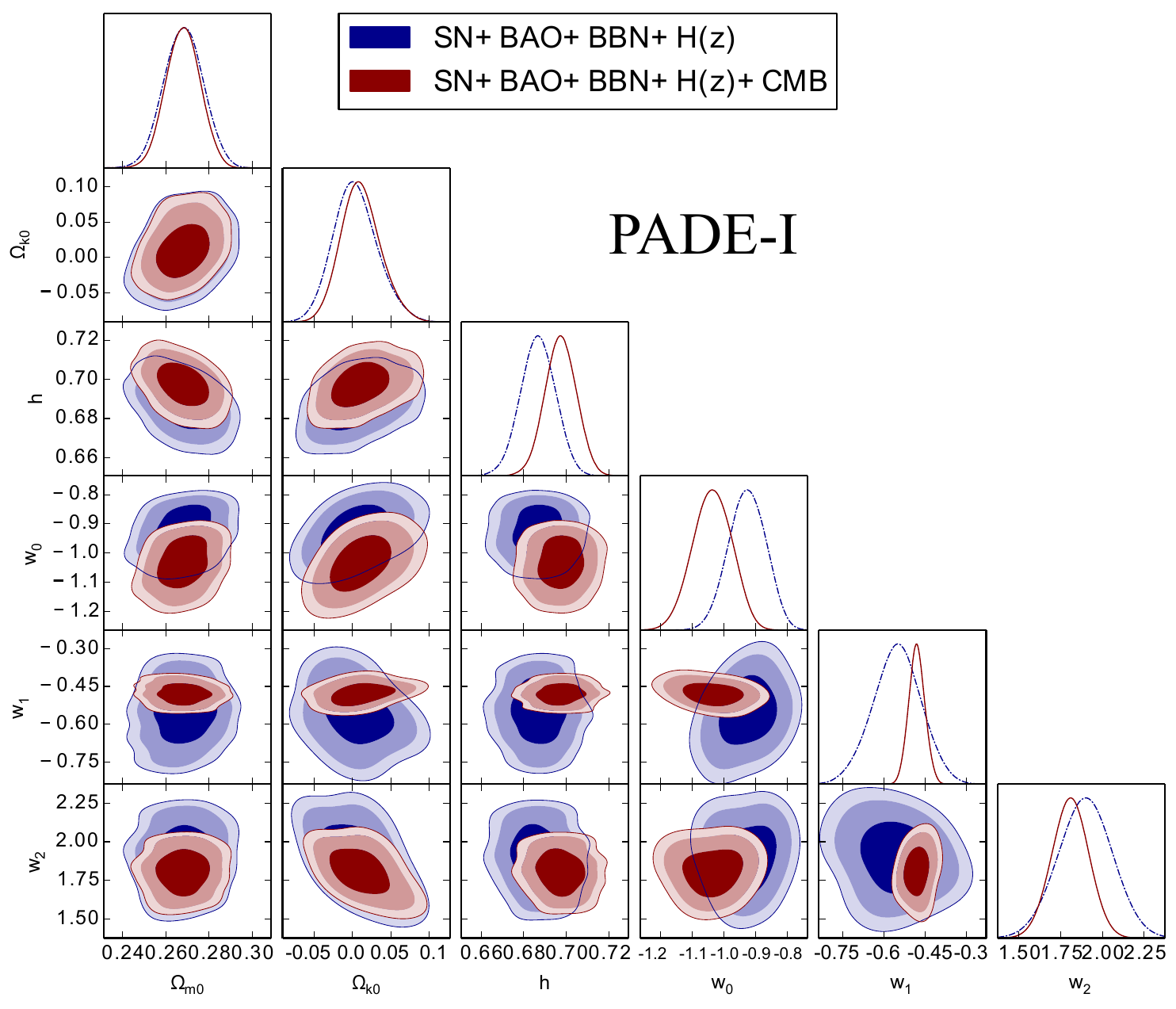}
\caption{One dimensional marginalized posterior distributions of the model parameters and the two-dimensional joint contours between several combinations of the model parameters at 68\% CL and 95\% CL of  the PADE-I parametrization for the combined datasets SN+BAO+BBN+$H(z)$ and SN+BAO+BBN+$H(z)$+CMB.  } 
    \label{fig:padeI}
\end{figure*}
\begin{figure*}
    \includegraphics[width=10cm]{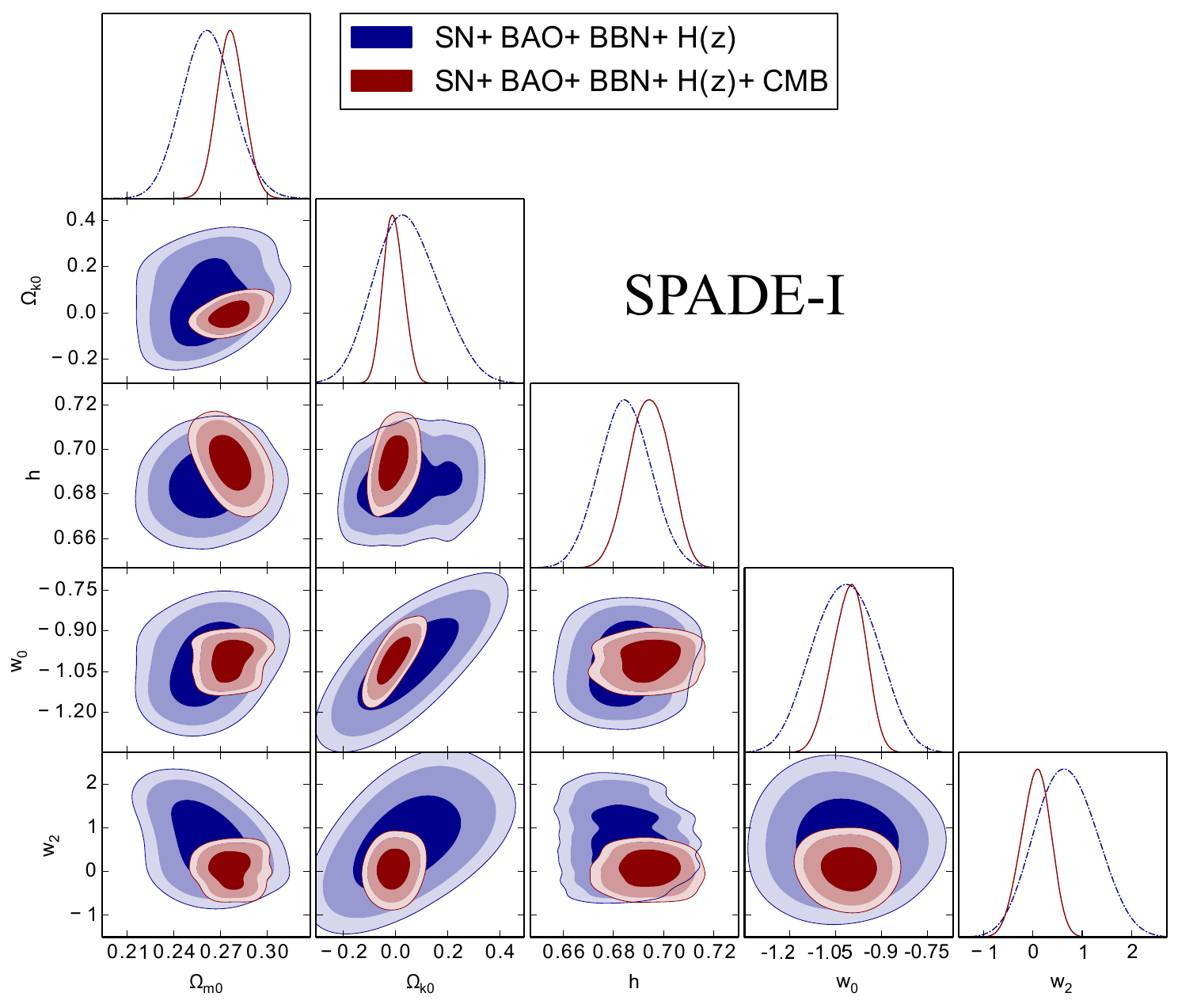}    
\caption{One dimensional marginalized posterior distributions of the model parameters and the two-dimensional joint contours between several combinations of the model parameters at 68\% CL and 95\% CL of  the SPADE-I parametrization for the combined datasets SN+BAO+BBN+$H(z)$ and SN+BAO+BBN+$H(z)$+CMB. } 
    \label{fig:spadeI}
\end{figure*}

\begin{figure*}
 \centering
    \includegraphics[width=10cm]{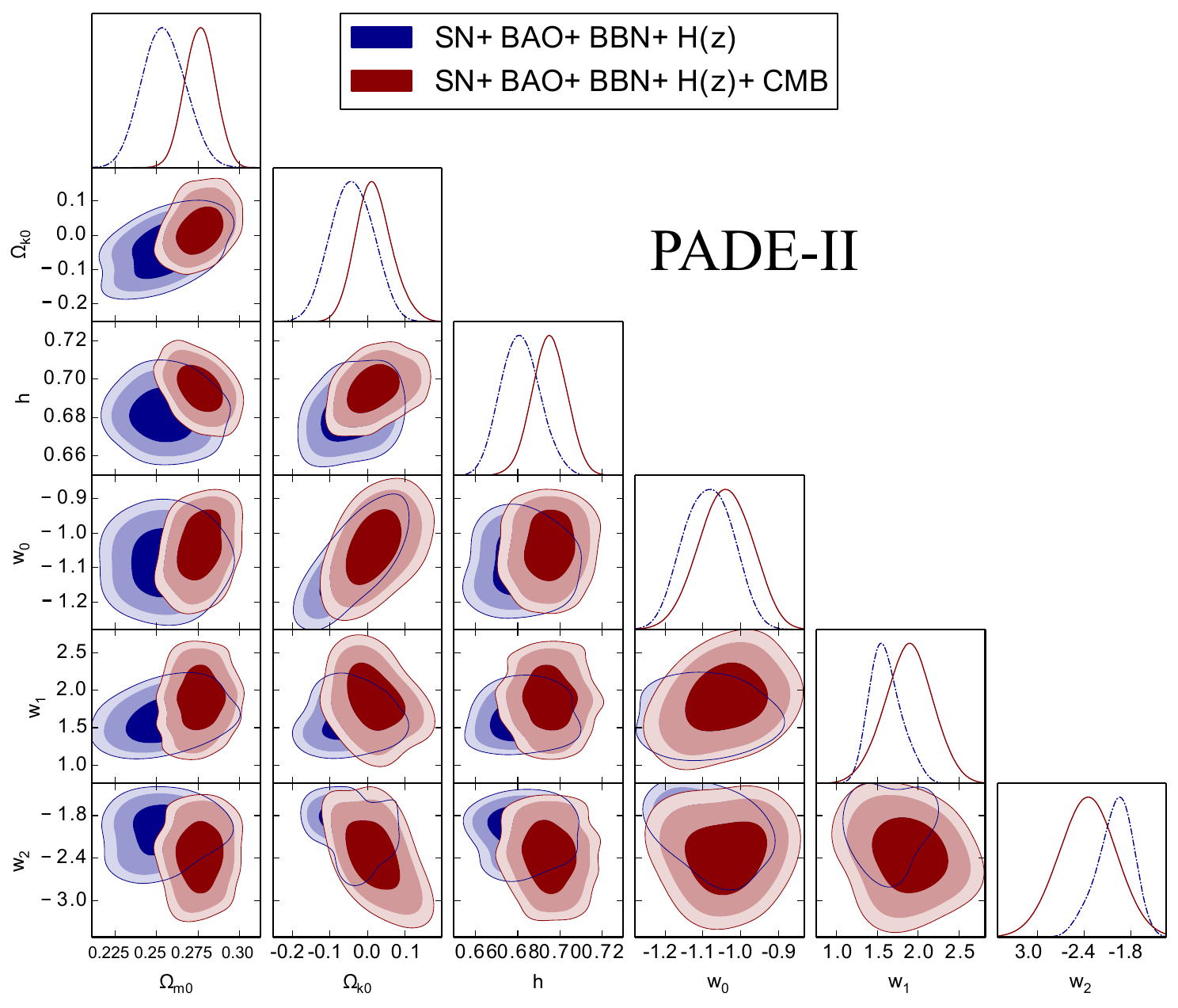}
\caption{One dimensional marginalized posterior distributions of the model parameters and the two-dimensional joint contours between several combinations of the model parameters at 68\% CL and 95\% CL of  the PADE-II parametrization for the combined datasets SN+BAO+BBN+$H(z)$ and SN+BAO+BBN+$H(z)$+CMB. } 
    \label{fig:padeII}
\end{figure*}
\begin{table*}
\centering
\caption{Summary of the 68\% CL constraints on various free parameters of the cosmological scenarios in the non-flat universe, namely, PADE-I, SPADE-I, PADE-II and $\Lambda$CDM for the combined datasets in presence of 1000 mock GWSS dataset, namely, SN+BAO+BBN+$H(z)$+GW and SN+BAO+BBN+$H(z)$+CMB+GW.  Note that $\Omega_{m0}$ denotes the total density parameter for the matter sector at present time, that means, $\Omega_{\rm m0} = \Omega_{\rm b0}+\Omega_{\rm dm0}$.  }
 \begin{tabular}{c  c  c c c c c }
 \hline 
 Model & $\Omega_{m0}$  & $\Omega_{k0}$  & $h$ [km/s/Mpc]  & $w_0$   & $w_1$ 
 & $w_2$\\
 \hline 
\multicolumn{7}{c}{SN+BAO+BBN+$H(z)$+GW} \\
 \hline 
PADE-I & $0.2561\pm 0.0078$ & $0.0999^{+0.036}_{-0.026}$ & $0.6829\pm 0.0056$  & $-1.097^{+0.030}_{-0.025}$ & $-0.409\pm 0.096$  & $0.97\pm 0.15$  \\
 
 SPADE-I & $0.259\pm 0.013$ & $-0.099\pm 0.043$ & $0.6818\pm 0.0099$  & $-1.081\pm 0.062$ & $-$  & $-0.12\pm 0.17$ \\
 
 PADE-II & $0.264\pm 0.011$ & $0.039\pm 0.023$ & $0.6871\pm 0.0099$  & $-1.038\pm 0.045$ & $0.231\pm 0.077$  & $-0.903^{+0.092}_{-0.12}$ \\
 
 $\Lambda$CDM & $0.2915^{+0.0091}_{-0.011}$ & $0.034^{+0.017}_{-0.022}$ & $0.6789\pm 0.0094$ & $-$ &  $-$ & $-$ \\
 \hline 
\multicolumn{7}{c}{SN+BAO+BBN+$H(z)$+CMB+GW} \\
 \hline 
PADE-I & $0.2630\pm 0.0058$ & $0.073\pm 0.023$ & $0.6719\pm 0.0052$  & $-1.099\pm 0.024$ & $-0.397\pm 0.053$  & $1.464^{+0.072}_{-0.092}$  \\
 
 SPADE-I & $0.2739\pm 0.0076$ & $-0.061\pm 0.025$ & $0.6889\pm 0.0083$  & $-1.105^{+0.047}_{-0.042}$ & $-$  & $0.108\pm 0.062$ \\
 
 PADE-II & $0.2728\pm 0.0096$ & $0.043\pm 0.023$ & $0.6889\pm 0.0087$  & $-1.094\pm 0.040$ & $-0.257\pm 0.071$  & $-0.876^{+0.072}_{-0.092}$ \\
 
 $\Lambda$CDM & $0.2846\pm 0.0075$ & $0.0171^{+0.0084}_{-0.014}$ & $0.6944\pm 0.0062$ & $-$ &  $-$ & $-$ \\
\hline 
\hline 
\end{tabular}\label{tab:best-GW}
\end{table*}
\begin{table*}
\centering
\caption{The 68\% CL constraints on various free parameters of the cosmological scenarios in the  flat universe, namely, PADE-I, SPADE-I, PADE-II and $\Lambda$CDM for the combined datasets in presence of 1000 mock GWSS dataset, namely, SN+BAO+BBN+$H(z)$+GW and SN+BAO+BBN+$H(z)$+CMB+GW. Note that $\Omega_{m0}$ denotes the total density parameter for the matter sector at present time, that means, $\Omega_{\rm m0} = \Omega_{\rm b0}+\Omega_{\rm dm0}$.  }
 \begin{tabular}{c  c  c  c c c }
 \hline 
 Model & $\Omega_{m0}$  & $h$ [km/s/Mpc]  & $w_0$   & $w_1$ 
 & $w_2$\\
 \hline 
\multicolumn{6}{c}{SN+BAO+BBN+$H(z)$+GW} \\
 \hline 

PADE-I & $0.257\pm 0.0069$ &  $0.684\pm 0.0061$  & $-1.086^{+0.023}_{-0.027}$ & $-0.571\pm 0.087$  & $1.03\pm 0.09$  \\
 
SPADE-I & $0.261\pm 0.011$ & $0.685\pm 0.008$  & $-1.075\pm 0.060$ & $-$  & $0.766\pm 0.021$ \\

PADE-II & $0.271\pm 0.013$ & $0.691\pm 0.0093$  & $-0.998\pm 0.029$ & $0.084\pm 0.029$  & $-0.159^{+0.036}_{-0.032}$ \\
 
$\Lambda$CDM & $0.292 \pm 0.009$  & $0.682\pm 0.0091$ & $-$ &  $-$ & $-$ \\
 \hline 
\multicolumn{6}{c}{SN+BAO+BBN+$H(z)$+CMB+GW} \\
 \hline 

PADE-I & $0.264\pm 0.0054$ & $0.673\pm 0.004$  & $-1.081\pm 0.013$ & $-0.406\pm 0.042$  & $1.37^{+0.07}_{-0.08}$  \\

SPADE-I & $0.272\pm 0.0064$ & $0.695\pm 0.009$  & $-1.079^{+0.037}_{-0.040}$ & $-$  & $0.493\pm 0.052$ \\

PADE-II & $0.281\pm 0.0089$ & $0.692\pm 0.007$  & $-1.009\pm 0.037$ & $0.023\pm 0.039$  & $-0.107^{+0.054}_{-0.061}$ \\
 
$\Lambda$CDM & $0.286\pm 0.008$ & $0.696\pm 0.007$ & $-$ &  $-$ & $-$ \\
\hline 
\hline 
\end{tabular}\label{tab:flat-GW}
\end{table*}

\begin{figure*}
    \includegraphics[width=10cm]{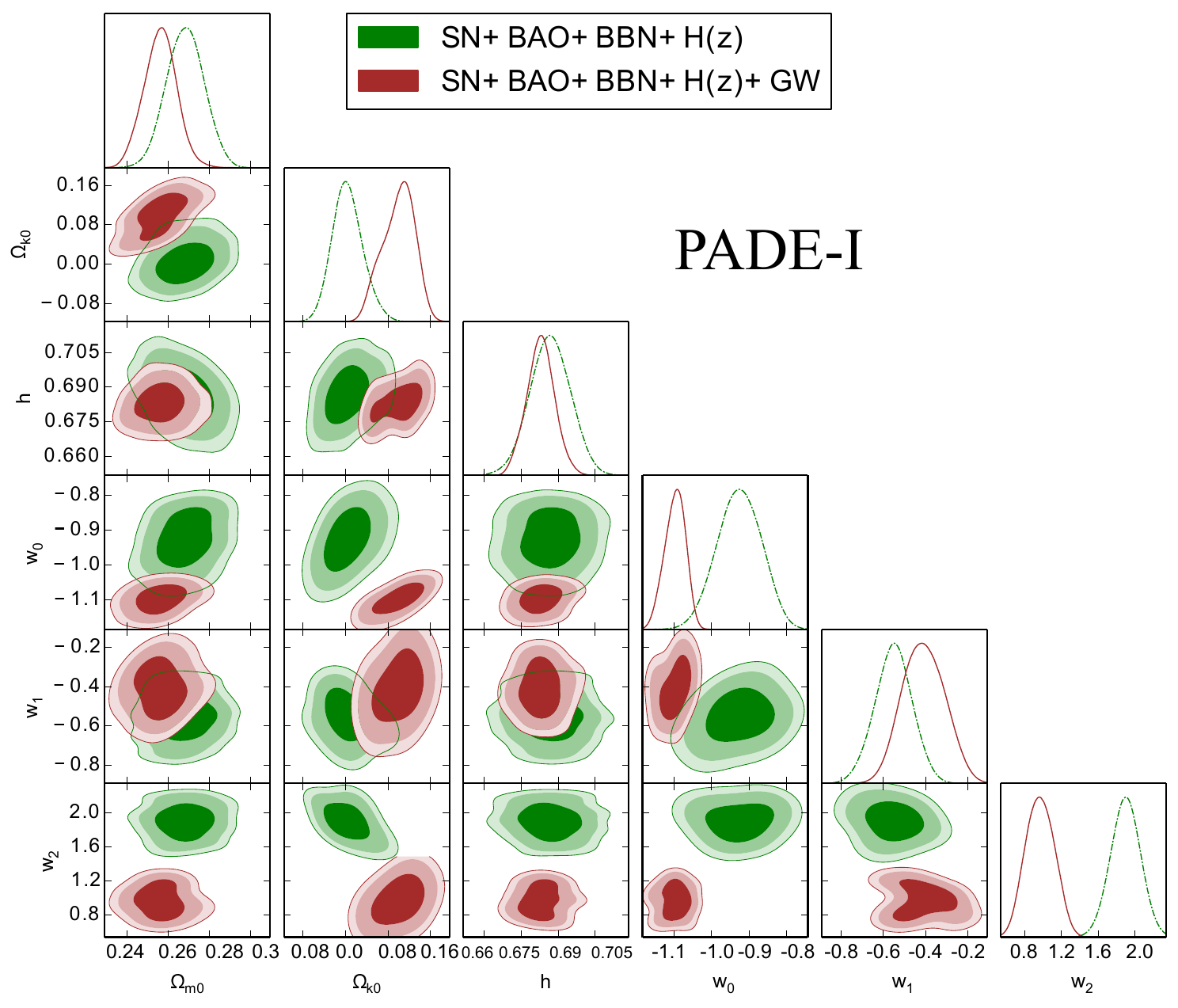}
\includegraphics[width=10cm]{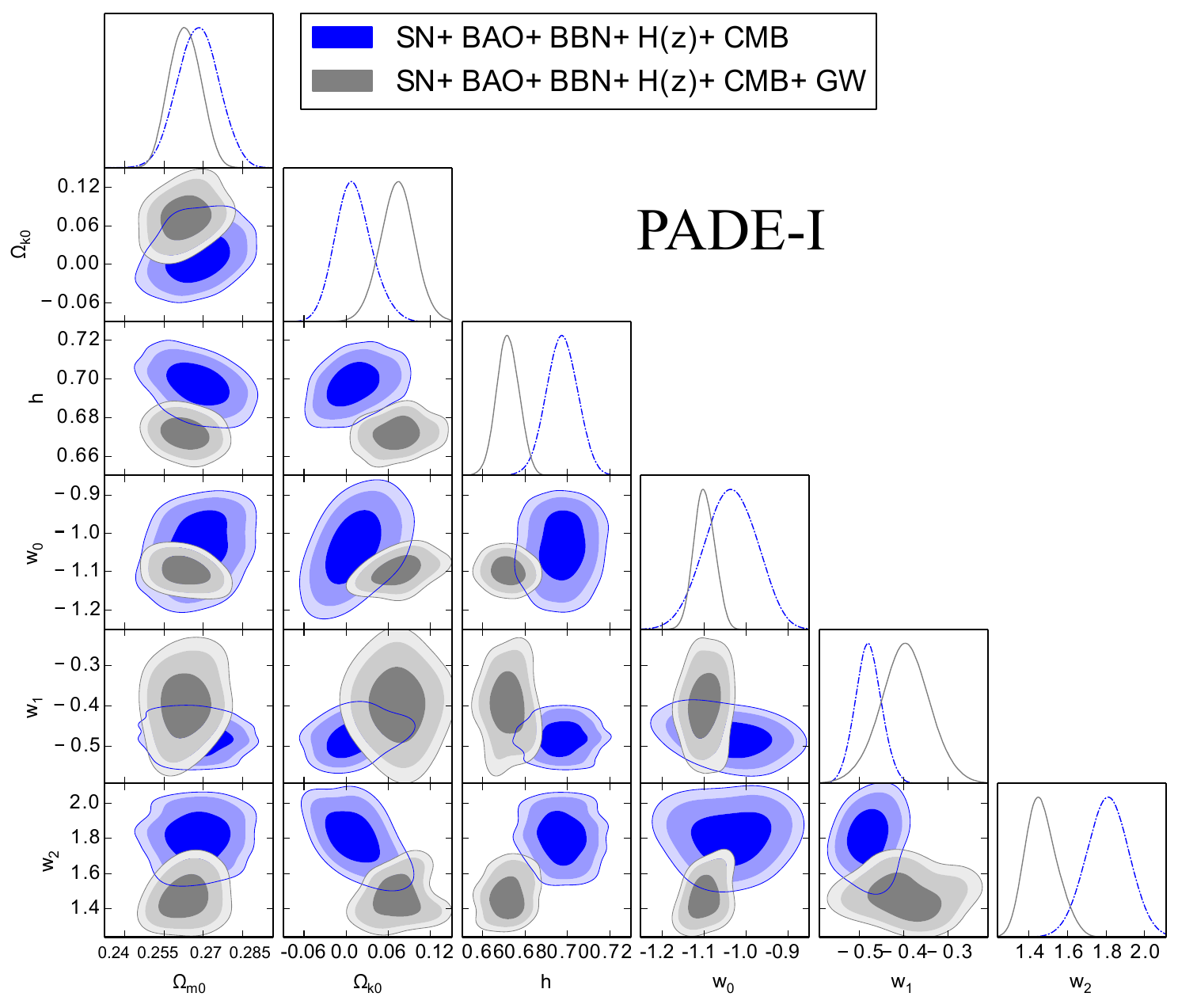}
\caption{The effects of adding 1000 mock GWSS luminosity distance measurements to the combined datasets SN+BAO+BBN+$H(z)$ and SN+BAO+BBN+$H(z)$+CMB have been shown through the one dimensional marginalized posterior distributions and the two-dimensional joint contours for the PADE-I parametrization.  } 
    \label{fig:padeI+gw}
\end{figure*}
\begin{figure*}
    \includegraphics[width=10cm]{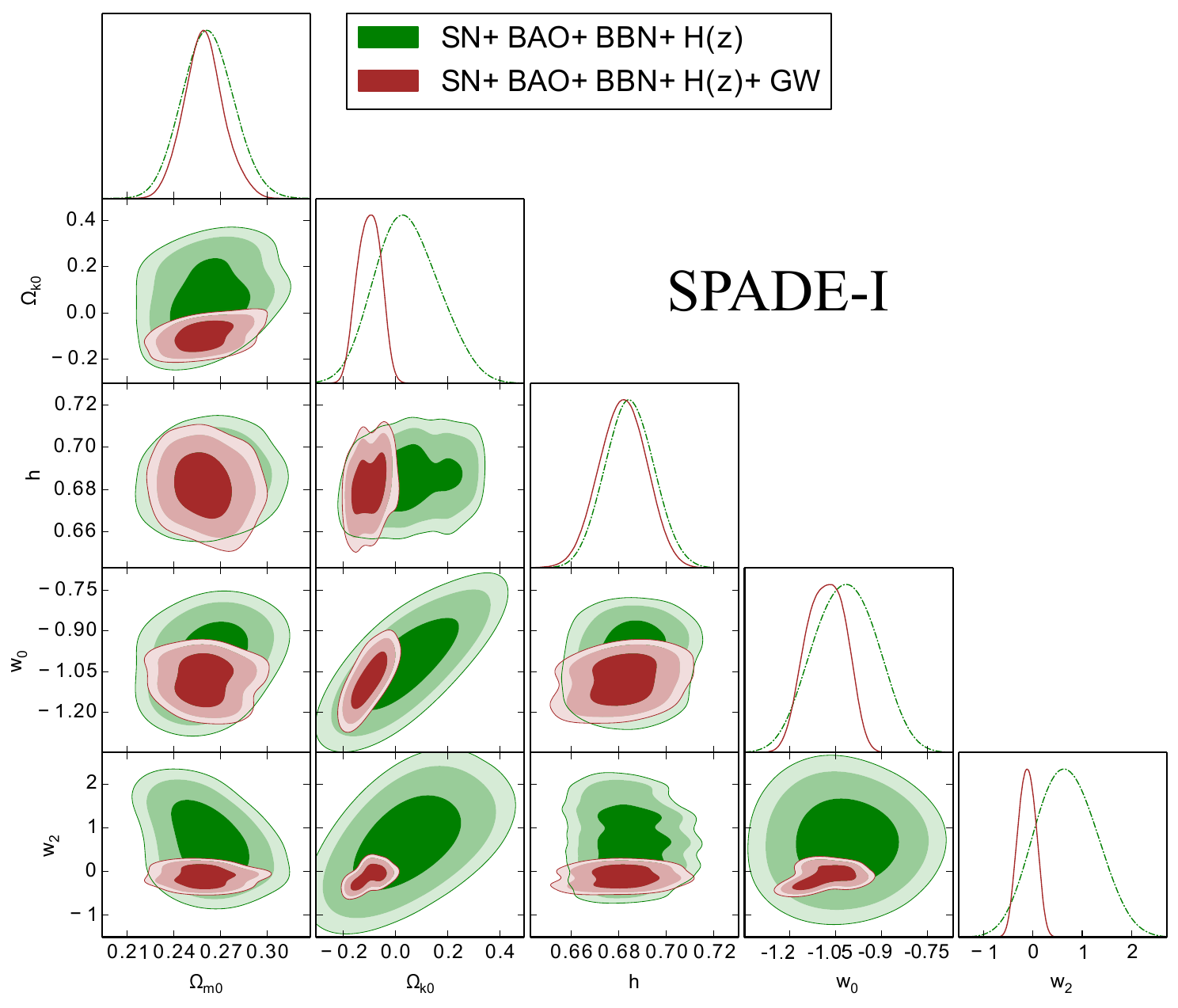}
    \includegraphics[width=10cm]{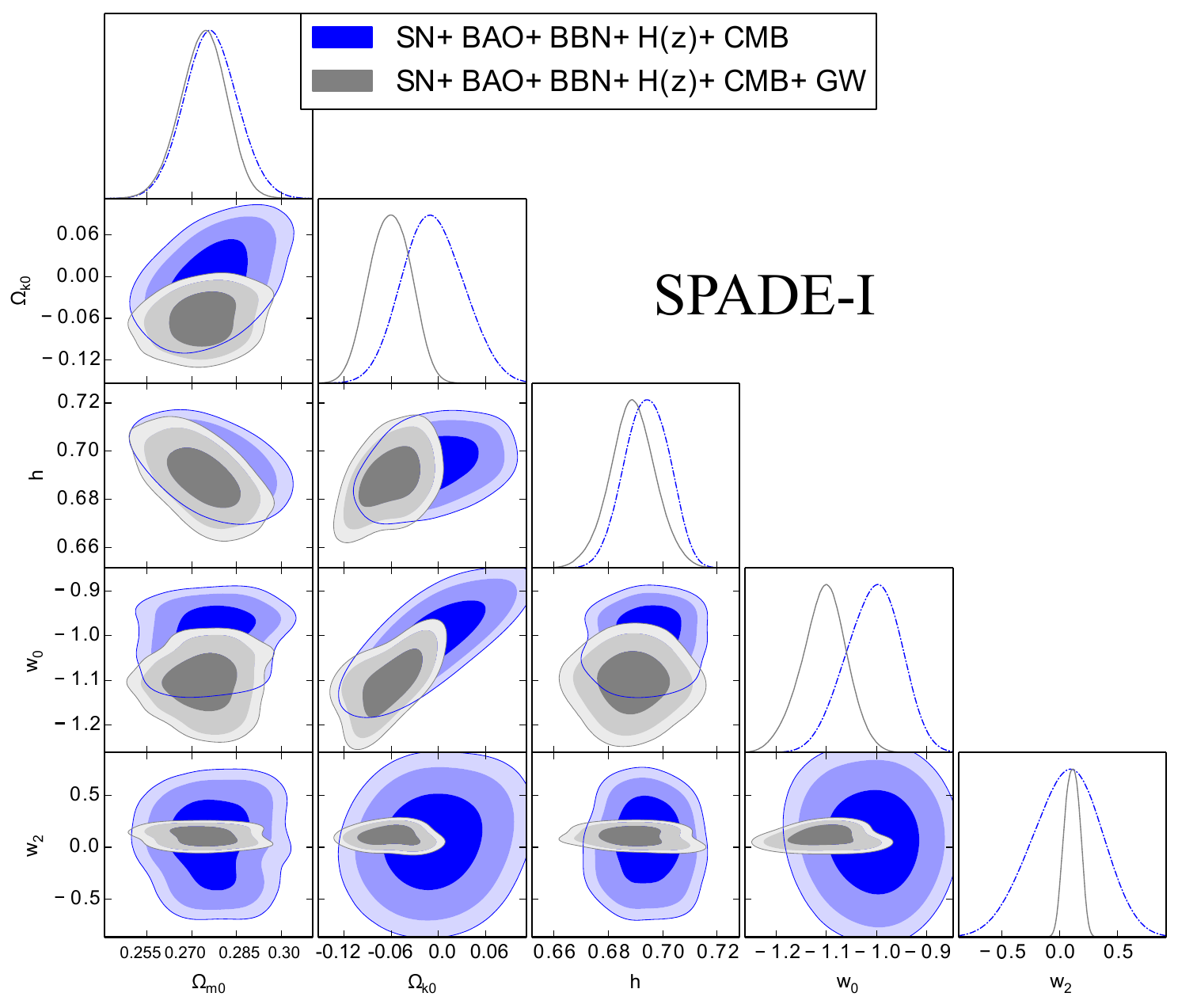}
\caption{The effects of adding 1000 mock GWSS luminosity distance measurements to the combined datasets SN+BAO+BBN+$H(z)$ and SN+BAO+BBN+$H(z)$+CMB have been shown through the one dimensional marginalized posterior distributions and the two-dimensional joint contours for the SPADE-I parametrization.   }
    \label{fig:spadeI+gw}
\end{figure*}
\begin{figure*}
    \includegraphics[width=10cm]{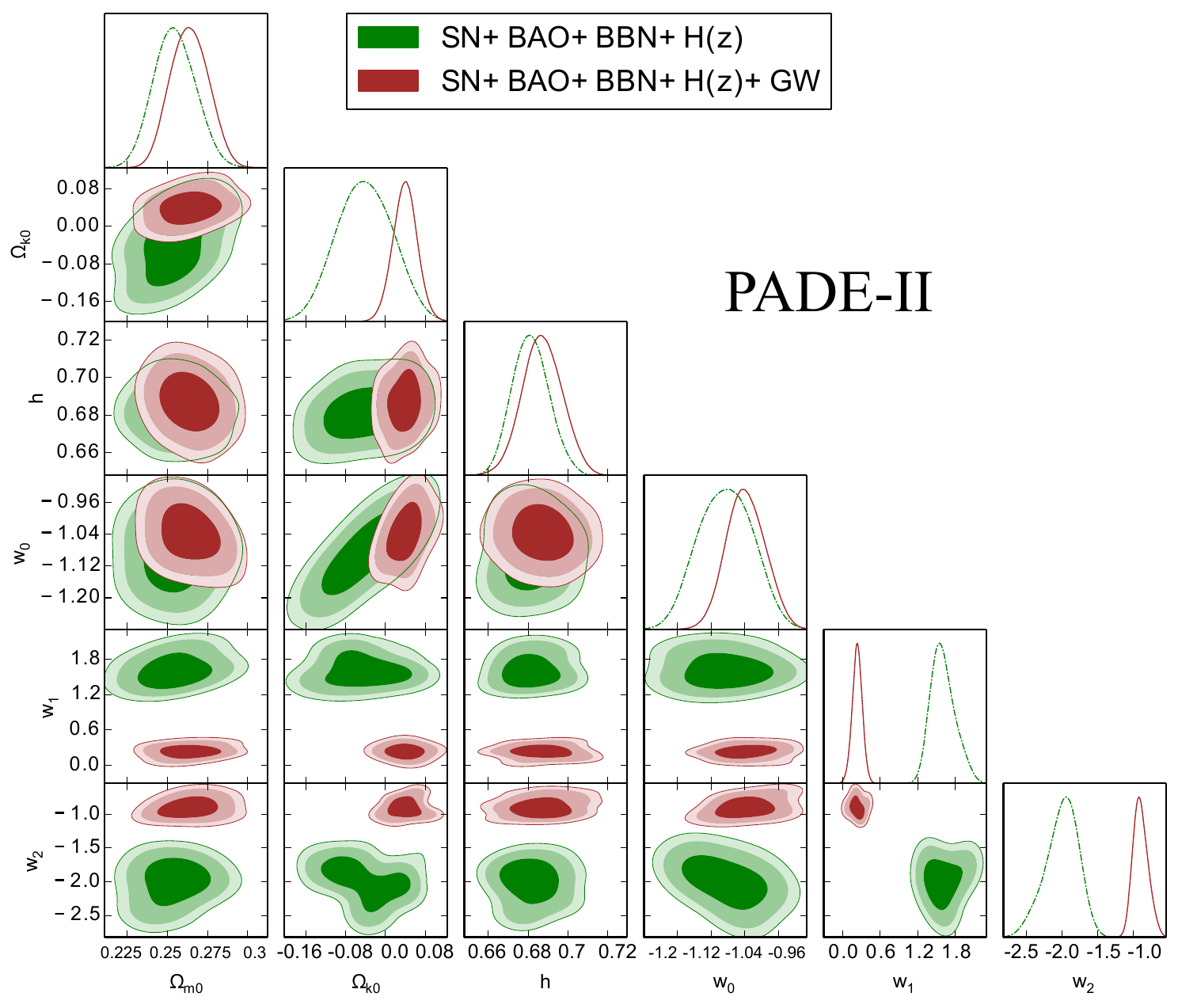}
    \includegraphics[width=10cm]{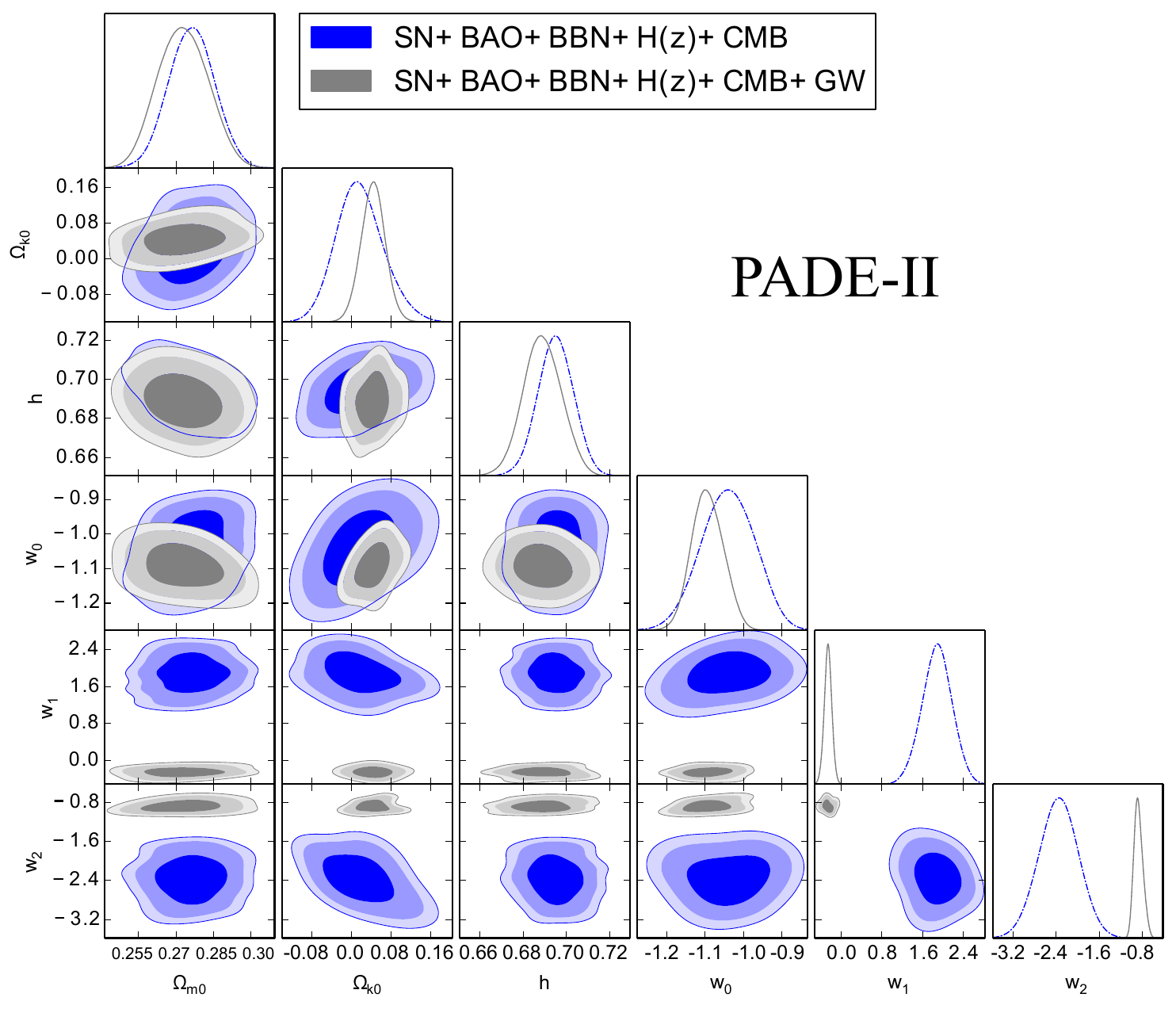}
    \caption{The effects of adding 1000 mock GWSS luminosity distance measurements to the combined datasets SN+BAO+BBN+$H(z)$ and SN+BAO+BBN+$H(z)$+CMB have been shown through the one dimensional marginalized posterior distributions and the two-dimensional joint contours for the PADE-II parametrization.}
    \label{fig:padeII+gw}
\end{figure*}
\begin{table*}
\caption{We summarize  $\chi^2_{\rm min}$, $\Delta \chi^2$~[$=$ $\chi^2_{\rm min}$ (PADE) $-$ $\chi^2_{\rm min}$ ($\Lambda$CDM)] and various information criteria for the cosmological scenarios  
PADE-I, SPADE-I, PADE-II with respect to the reference model $\Lambda$CDM in the context of a non-flat universe, obtained for the combined datasets SN+BAO+BBN+$H(z)$, SN+BAO+BBN+$H(z)$+CMB, SN+BAO+BBN+$H(z)$+GW and SN+BAO+BBN+$H(z)$+CMB+GW. Here $K$ denotes the number of free parameters of the cosmological models considered in the statistical data analysis, and $N$ is the total number of data points in each combined dataset. Following the definition of $\Delta \chi^2$, one can see that $\Delta \chi^2 <0$ implies the preference of the PADE model over the $\Lambda$CDM and $\Delta \chi^2 >0$ dictates the opposite.  }
 \begin{tabular}{c c c c c c c c c c c}
 \hline 
 Model &~ $\chi^2_{\rm min}$ &~ $\Delta \chi^2$  &~ $K$  &~ $N$  &~  AIC &~ $\Delta$AIC &~ BIC &~ $\Delta$BIC &~ DIC &~ $\Delta$DIC\\
 \hline 
 \multicolumn{10}{c}{SN+BAO+BBN+$H(z)$} \\
 \hline 
 
PADE-I & 1056.0 & $-3.3$ & 7 & 1092 & 1070.0 & 2.7 & 1105.0 & 17.7 & 1068.6 & 0.5\\
 
SPADE-I & 1057.2 & $-2.1$ & 6 & 1092 & 1069.2 & 1.9 & 1099.2 & 11.9 & 1068.6 & 0.5\\
 
PADE-II & 1056.7 & $-2.6$ & 7 & 1092 & 1070.7 & 3.4 & 1105.7 & 18.4 & 1068.9 & 0.8\\
 
$\Lambda$CDM &  1059.3 & $-$ & 4 & 1092 & 1067.3 & $-$ & 1087.3 & $-$ & 1068.1  & $-$ \\
 \hline 
 
 \multicolumn{10}{c}{SN+BAO+BBN+$H(z)$+CMB} \\
 \hline 
PADE-I & 1060.2 & $-1.1$ & 7 & 1095 & 1074.2 & 4.9 & 1109.2 & 19.9 & 1072.4 & 1.1\\
 
SPADE-I & 1061.4 & $0.1$ & 6 & 1095 & 1073.4 & 4.1 & 1103.4 & 14.1 & 1072.1 & 0.8\\
 
PADE-II & 1060.0 & $-1.3$ & 7 & 1095 & 1074.0 & 4.7 & 1109.0 & 19.7 & 1072.7 & 1.4\\
 
$\Lambda$CDM & 1061.3 & $-$ & 4 & 1095 & 1069.3 & $-$ & 1089.3 & $-$ & 1071.3  & $-$  \\

 \hline 
 \multicolumn{10}{c}{SN+BAO+BBN+$H(z)$+GW} \\
 \hline 
 
PADE-I & 2058.9 & $-3.5$ &  7 & 2092 & 2072.9 & 2.5 & 2107.9 & 17.5 & 2070.1 & 0.7\\
 
SPADE-I & 2060.0 & $-2.4$ &  6 & 2092 & 2072.0 & 1.6 & 2102.0 & 11.6 & 2069.6 & 0.2\\
 
PADE-II & 2059.1 &  $-3.3$ &  7 & 2092 & 2073.1 & 2.7 & 2108.1 & 17.7 & 2069.7 & 0.3\\
 
$\Lambda$CDM & 2062.4  & $-$ & 4 & 2092 & 2070.4 & $-$ & 2090.4 & $-$ & 2069.4  & $-$\\
 \hline 
 
 \multicolumn{10}{c}{SN+BAO+BBN+$H(z)$+CMB+GW} \\
 \hline 
PADE-I & 2063.4 & $-2.2$ &  7 & 2095 & 2077.4 & 3.8 & 2112.4 & 18.8 & 2073.8 & 0.8\\
 
SPADE-I & 2062.5 & $-3.1$ &  6 & 2095 & 2074.5 & 0.9 & 2104.5 & 10.9 & 2072.9 & -0.1\\
 
PADE-II & 2062.3 & $-3.3$ &  7 & 2095 & 2076.3 & 2.7 & 2111.3 & 17.7 & 2073.1 & 0.1\\
 
$\Lambda$CDM & 2065.6 & $-$  & 4 & 2095 & 2073.6 & $-$ & 2093.6 & $-$ & 2073.0  & $-$\\
  
\hline 

\end{tabular}\label{tab:IC}
\end{table*}

\begin{figure*}
    \centering
    \includegraphics[width=5cm]{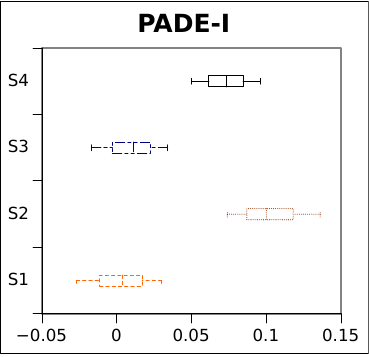}
    \includegraphics[width=5cm]{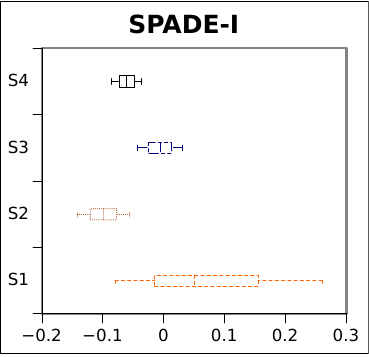}
    \includegraphics[width=5cm]{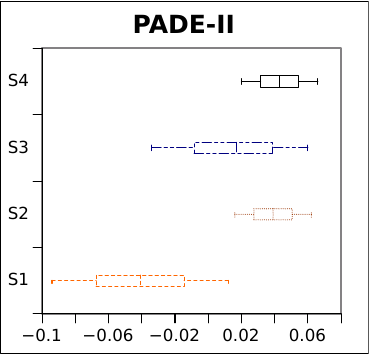}
    \caption{Whisker plots with  68\% and 95\% CL constraints on $\Omega_{k0}$ for all three parametrizations have been shown for all the combined datasets, namely  SN+BAO+BBN+$H(z)$ (labeled as {\bf S1}), SN+BAO+BBN+$H(z)$+GW (labeled as {\bf S2}), SN+BAO+BBN+$H(z)$+CMB (labeled as {\bf S3}) and SN+BAO+BBN+$H(z)$+CMB+GW (labeled as {\bf S4}). }
    \label{fig:OMEGAK}
\end{figure*}
\begin{figure*}
    \centering
    \includegraphics[width=5cm]{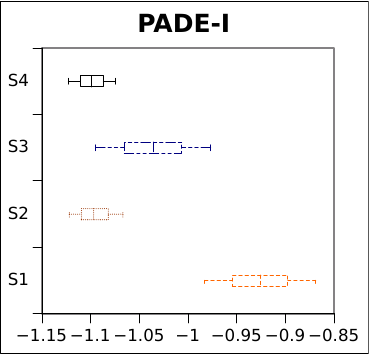}
    \includegraphics[width=5cm]{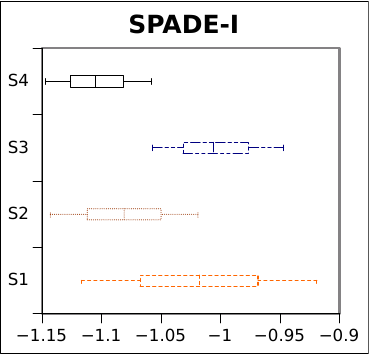}
    \includegraphics[width=5cm]{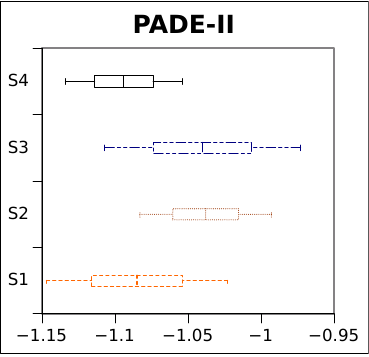}
    \caption{Whisker plots with  68\% and 95\% CL constraints on $w_0$ for all three parametrizations have been shown for all the combined datasets, namely  SN+BAO+BBN+$H(z)$ (labeled as {\bf S1}), SN+BAO+BBN+$H(z)$+GW (labeled as {\bf S2}), SN+BAO+BBN+$H(z)$+CMB (labeled as {\bf S3}) and SN+BAO+BBN+$H(z)$+CMB+GW (labeled as {\bf S4}). }
    \label{fig:w0}
\end{figure*}

\begin{figure*}
    \centering
    \includegraphics[width=5cm]{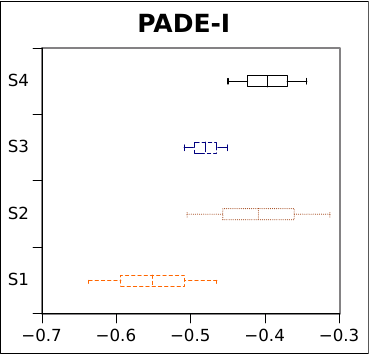}
    \includegraphics[width=5cm]{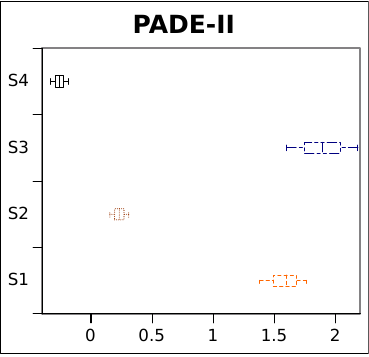}
    \caption{Whisker plots with  68\% and 95\% CL constraints on $w_1$ for all three parametrizations have been shown for all the combined datasets, namely  SN+BAO+BBN+$H(z)$ (labeled as {\bf S1}), SN+BAO+BBN+$H(z)$+GW (labeled as {\bf S2}), SN+BAO+BBN+$H(z)$+CMB (labeled as {\bf S3}) and SN+BAO+BBN+$H(z)$+CMB+GW (labeled as {\bf S4}).}
    \label{fig:w1}
\end{figure*}

\begin{figure*}
    \centering
    \includegraphics[width=5cm]{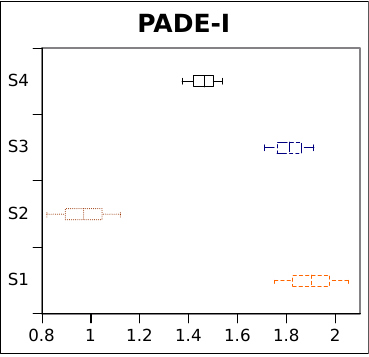}
    \includegraphics[width=5cm]{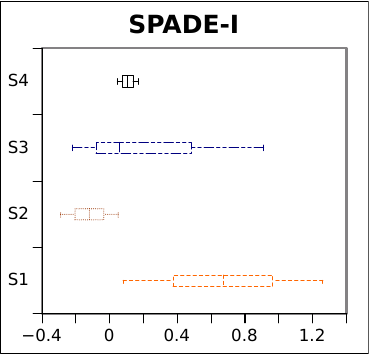}
    \includegraphics[width=5cm]{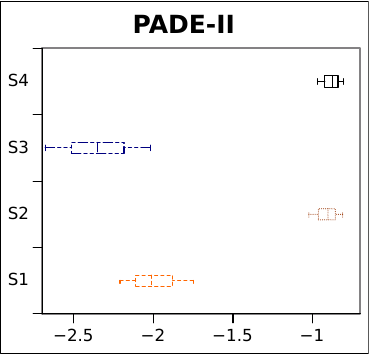}
    \caption{Whisker plots with  68\% and 95\% CL constraints on $w_2$ for all three parametrizations have been shown for all the combined datasets, namely  SN+BAO+BBN+$H(z)$ (labeled as {\bf S1}), SN+BAO+BBN+$H(z)$+GW (labeled as {\bf S2}), SN+BAO+BBN+$H(z)$+CMB (labeled as {\bf S3}) and SN+BAO+BBN+$H(z)$+CMB+GW (labeled as {\bf S4}).}
    \label{fig:w2}
\end{figure*}

\section{Results and Analyses}
\label{sec-results}

In this section we describe the constraints on the proposed PADE scenarios considering initially the curvature density parameter as a free parameter, and then we extract the constraints on the same scenarios setting the curvature density parameter to zero, that means considering the flat universe in the background. Along with the PADE scenarios, we have considered the $\Lambda$CDM cosmological models as the reference models. That means, for the PADE scenarios in the non-flat background, our reference model is the non-flat $\Lambda$CDM model and for the PADE scenarios in the flat universe, our reference model is the flat $\Lambda$CDM model. We have presented the constraints taking the combined analyses of the usual cosmological probes, e.g. SN+BAO+BBN+$H(z)$, SN+BAO+BBN+$H(z)$+CMB, and then we have included the GWSS mock dataset with these combined probes, i.e. the combined datasets SN+BAO+BBN+$H(z)$+GW and SN+BAO+BBN+$H(z)$+CMB+GW. Section \ref{sec-results-usual-probes} contains the results for the usual cosmological probes and  section~\ref{sec-results-gwss} presents the results for the mock datasets. Finally,  in section \ref{sec-model-comparison} we have considered various information criteria in order to assess the models. The results are shown in 
Tables~\ref{tab:best}, \ref{tab:flat1}, \ref{table-H0}, \ref{tab:best-GW}, \ref{tab:flat-GW}, and \ref{tab:IC}. The plots displaying various parameters of the non-flat PADE parametrizations are Figs. \ref{fig:padeI}, \ref{fig:spadeI}, \ref{fig:padeII}, \ref{fig:padeI+gw}, \ref{fig:spadeI+gw}, \ref{fig:padeII+gw}, \ref{fig:OMEGAK}, \ref{fig:w0}, \ref{fig:w1} and \ref{fig:w2}. 

\subsection{Constraints from the current cosmological probes}
\label{sec-results-usual-probes}

Table \ref{tab:best} presents the 68\% constraints on the model parameters of PADE-I, SPADE-I and PADE-II and $\Lambda$CDM considering the non-flat universe for SN+BAO+BBN+$H(z)$ and SN+BAO+BBN+$H(z)$+CMB. The graphical distributions of the cosmological scenarios in the non-flat universe are displayed in Figs. \ref{fig:padeI}, \ref{fig:spadeI}, \ref{fig:padeII}, where we show the one dimensional marginalized posterior distributions of the model parameters and the two dimensional joint contours between several combinations of the model parameters. Additionally, in Table~\ref{tab:flat1}, we present the constraints on the parameters of the same cosmological scenarios but assuming the spatially flat universe.  We begin with the results obtained from the combined dataset SN+BAO+BBN+$H(z)$ (Table~\ref{tab:best}). Focusing on the curvature parameter, we notice that for all the DE parametrizations,  $\Omega_{k0}$ assumes very small values which allow its non-value within 68\% CL,  and hence the flat universe scenario is consistent within these DE parametrizations. 
The inclusion of the curvature parameter, as we observe, may affect the present cosmological parameters. This is seen from Figs. \ref{fig:padeI}, \ref{fig:spadeI}, \ref{fig:padeII} where we notice the correlation or anti-correlation between $\Omega_{k0}$ and other parameters. The free parameters of the DE parametrizations are thus equally affected due to the presence of the curvature parameter. Concerning the dark energy equation of state at present time, we find that for this combined dataset, i.e. SN+BAO+BBN+$H(z)$,  
$w_0 = -0.926\pm 0.057$ (for PADE-I), $-1.018\pm 0.099$ (for SPADE-I), $-1.085\pm 0.062$ (for PADE-II). Thus,  we conclude that for PADE-II, $w_0$ remains in the phantom regime at more than 68\% CL, while for the remaining two cases, $w_0$ allows both quintessence ($w_0> -1$) and phantom ($w_0 < -1$) values within 68\% CL. Now focusing on the remaining two DE parameters, $w_1$ and $w_2$ which quantify the dynamical nature in the DE parametrization (note that SPADE-I has only two free parameters, $w_0$ and $w_2$ unlike other two parametrizations having three free parameters $w_0$, $w_1$, $w_2$), we find that for PADE-I and PADE-II, both $w_1$ and $w_2$ are non-zero at more than 99\% CL (see Figs. \ref{fig:padeI} and \ref{fig:padeII} which clearly confirm this observation) with an improvement of $\Delta \chi^2 =  -3.3$ for PADE-I and $\Delta \chi^2 = -2.6$ for PADE-II, respectively. 
Hence, a dynamical nature in both these parametrizations is  strongly supported by this combined dataset. For SPADE-I, we observe that $w_2 \neq 0$ at more than 68\% CL, but it is consistent to zero within 95\% CL with an overall improvement of $\Delta \chi^2 = -2.1$. Thus, its dynamical nature is retained within 68\% CL. 
 Now, comparing the results from the same combined dataset (i.e. SN+BAO+BBN+$H(z)$), but for the spatially flat case (Table~\ref{tab:flat1}), we notice that $w_0$ for all three parametrizations has a mild indication towards the phantom region but they are consistent with $w_0  =-1$ within 68\% CL (similar conclusion has been found in Ref. \cite{Escamilla:2023oce}). Regarding the dynamical nature of the parametrizations quantified through $w_1$ and $w_2$,  our results remain same similar to what we observed in the non-flat case except some changes in the constraints in the flat case and such changes are due to the exclusion of $\Omega_{k0}$ from the picture. For example, in flat PADE-II scenario, we observe that $w_1$ is consistent to zero within 68\% CL but $w_2$ remains non-zero at more than 95\% CL, and hence, the dynamical nature of DE within flat PADE-II does not alter. 

Next we move to the results for the combined dataset
SN+BAO+BBN+$H(z)$+CMB in the non-flat universe which are displayed in the last half of Table~\ref{tab:best}. Looking at the constraints on all the parameters, one can clearly notice that the inclusion of CMB to SN+BAO+BBN+$H(z)$ affects the parameter space significantly. The error bars on almost all the parameters are reduced due to the inclusion of CMB to the former dataset SN+BAO+BBN+$H(z)$ but alongside the changes in the mean values in some parameters are also observed. 
Looking further at Figs. \ref{fig:padeI}, \ref{fig:spadeI}, \ref{fig:padeII}, we argue that the reduction in the parameter space depends on the  underlying cosmological scenario and this improvement is not generally true for all models. We first focus on the constraints on curvature parameter for SN+BAO+BBN+$H(z)$+CMB. Similar to the earlier constraints achieved from SN+BAO+BBN+$H(z)$, here too, we do not find any strong evidence of $\Omega_{k0} \neq 0$ in any of these parametrizations. We notice that $\Omega_{k0}$ assumes small values for all the parametrizations, and within 68\% CL, $\Omega_{k0} =0$ is allowed irrespective of the parametrizations.  Further, due to the presence of the correlation (anti-correlation) between $\Omega_{k0}$ and other free parameters including $w_i$'s, $i =0, 1,2$ (see Figs. \ref{fig:padeI}, \ref{fig:spadeI}, \ref{fig:padeII})\footnote{Let us note that for PADE-I, the existing anti-correlation between $\Omega_{k0}$ and $w_1$ for the dataset SN+BAO+BBN+$H(z)$ disappears for SN+BAO+BBN+$H(z)$+CMB (see Fig. \ref{fig:padeI}), however, for PADE-II, this does not happen (see Fig.~\ref{fig:padeII}). }, the  nature and dynamics of DE are also affected compared to the zero curvature case.  We find that $w_0$, the present value of the DE equation of state has a tendency of lying in the phantom regime for all the parametrizations: $w_0 = -1.036\pm 0.059$ (68\% CL, PADE-I), $-1.006^{+0.058}_{-0.051}$ (68\% CL, SPADE$-$I), $-1.040\pm 0.067$ (68\% CL, PADE-II). However, $w_0$ is consistent with the cosmological constant in all these scenarios.  The inclination of DE equation of state towards the phantom regime (assuming a curvature free universe in the background) has also been pointed out recently  with cosmological constant being the best solution \cite{Escamilla:2023oce}.   Concerning the remaining two parameters, $w_1$ and $w_2$, we find that for both PADE-I and PADE-II, $w_1 \neq 0$ and $w_2 \neq 0$ at more than 99\% CL  with an improvement of $\Delta \chi^2 = -1.1$ (for PADE-I) and $\Delta \chi^2 = -1.3$ (for PADE-II), see Table~\ref{tab:IC}. We again refer to Figs. \ref{fig:padeI} and \ref{fig:padeII} which prove this clearly. 
Thus, we infer that the combined dataset SN+BAO+BBN+$H(z)$+CMB indicates the dynamical nature in the DE equation-of-state for these two parametrizations. This is an interesting result of this article which goes in an agreement with Ref. \cite{Zhao:2017cud} indicating the dynamical nature in the DE fluid as an effect of the current cosmological tensions. 
For the remaining parameter $w_2$ in SPADE$-$I, we do not find any evidence of its non-zero nature ($w_2 = 0.06^{+0.31}_{-0.28}$ at 68\% CL) and thus,  we do not find any strong preference for a dynamical DE within this parametrization. 
Now, comparing the findings on $w_1$, $w_2$ in the flat universe (see Table~\ref{tab:flat1}), we notice that their constraints are significantly changed than in the non-flat universe (Table~\ref{tab:best}). For example, within the flat SPADE-I parametrization, we find evidence of $w_2 \neq 0$ at more than several standard deviations which was absent in the non-flat case. However, for the remaining two cases,  we find evidence of $(w_1, w_2) \neq (0, 0)$, that means, dynamical nature of DE does not alter but it is weakened because of the present constraints in the flat space.

Finally, we concentrate on the Hubble constant estimations within these DE parametrizations considering both the combined analyses SN+BAO+BBN+$H(z)$ and SN+BAO+BBN+$H(z)$+CMB. From Table~\ref{tab:best} we see that for both SN+BAO+BBN+$H(z)$ and SN+BAO+BBN+$H(z)$+CMB, the estimated values of $h~(=H_0/100)$ within these PADE parametrizations are similar to the non-flat $\Lambda$CDM model. However, the only difference is that, for SN+BAO+BBN+$H(z)$+CMB, $h$ assumes mildly higher values compared to its estimations from SN+BAO+BBN+$H(z)$. The constraints on $h$ also remain similar when we consider the flat case (see Table~\ref{tab:flat1}). 
Nevertheless, all of them are still far from the SH0ES (Supernovae and $H_0$ for the Equation of State of DE) collaboration \cite{Riess:2021jrx}. In order to understand how the recent SH0ES prior ($H_0 = 73.04 \pm 1.04 $ km/s/Mpc at 68\% CL) \cite{Riess:2021jrx} may affect the constraints on the reduced Hubble constant $h$ within these cosmological scenarios, we added the said $H_0$ prior to both SN+BAO+BBN+$H(z)$ and SN+BAO+BBN+$H(z)$+CMB considering both the flat and nonflat cases separately. The results are summarized in Table~\ref{table-H0} where a comparison between the values of $h$ obtained in different combined analyses are shown.  One can clearly see from Table~\ref{table-H0} that the $H_0$ prior indeed affects the constraints on $h$ leading to higher values.   


\subsection{Constraints in presence of the GWSS luminosity distance measurements}
\label{sec-results-gwss}

Here we present the constraints on the proposed parametrizations when the future GWSS dataset is added to the usual cosmological probes discussed in section~\ref{sec-results-usual-probes}. In order to understand how the future GWSS data
could affect the constraints on the present cosmological scenarios, we have combined the 1000 mock GWSS luminosity distance measurements with SN+BAO+BBN+$H(z)$ and SN+BAO+BBN+$H(z)$+CMB. In Table~\ref{tab:best-GW}, we have presented the results on all the scenarios (including the non-flat $\Lambda$CDM) considering two combined datasets, namely, SN+BAO+BBN+$H(z)$+GW and SN+BAO+BBN+$H(z)$+CMB+GW.  Additionally, in Table~\ref{tab:flat-GW}, we present the constraints on the parameters of the same cosmological scenarios but assuming the spatially flat universe in the background.  
In Figs. \ref{fig:padeI+gw}, \ref{fig:spadeI+gw}, \ref{fig:padeII+gw} we have graphically shown the effects of mock GWSS on the cosmological parameters of the three distinct cosmological scenarios when combined with SN+BAO+BBN+$H(z)$ and SN+BAO+BBN+$H(z)$+CMB. 
Before presenting the main results, we notice that for PADE-I, corresponding to the upper plot in Fig. \ref{fig:padeI+gw}, the parameter $w_1$ is found to be in tension at more than $3\sigma$ between SN+BAO+BBN+$H(z)$ and SN+BAO+BBN+$H(z)$+GW.  However, this vanishes when we compared the results between SN+BAO+BBN+$H(z)$+CMB and SN+BAO+BBN+$H(z)$+CMB+GW (see the lower plot in Fig. \ref{fig:padeI+gw}). For SPADE-I, corresponding to Fig. \ref{fig:spadeI+gw}, we do not find any tension in any parameter for all the datasets.  For PADE-II, the DE parameters $w_1$ and $w_2$ are found to be in tension at more than $3\sigma$ for all the datasets. This might be caused either due to a large number of free parameters in the parametrizations or because of not enough sensitivity of the mock GWSS dataset.  Nevertheless,  here we present the impact of the 1000 mock GWSS dataset when it is combined with the usual cosmological probes.

Now, looking at the results obtained from SN+BAO+BBN+$H(z)$ (Table~\ref{tab:best}) and SN+BAO+BBN+$H(z)$+GW (Table~\ref{tab:best-GW}), one can clearly see that  the mock GWSS dataset could affect the cosmological parameters significantly. For instance, focusing on the curvature density parameter, we observe that when the GWSS dataset is added to  SN+BAO+BBN+$H(z)$, $\Omega_{k0}$ takes non-zero values at more than 68\% CL for all the scenarios. On the other hand, the reduction in the uncertainties in $\Omega_{k0}$ for both SPADE-I and PADE-II is clear from Table~\ref{tab:best-GW}. Concentrating on the present value of the DE equation of state, $w_0$, we see that the mean values of $w_0$ in all three PADE scenarios are directed towards the phantom regime when we consider  SN+BAO+BBN+$H(z)$+GW (Table~\ref{tab:best-GW}). In particular, $w_0 < -1$ at more than 68\% CL for both PADE-I and SPADE-I (for SN+BAO+BBN+$H(z)$+GW) while for PADE-II, $w_0 =-1$ is consistent within 68\% CL for the same dataset. 
Regarding the $w_1$ parameter for both PADE-I and PADE-II, we find that the mean values of $w_1$ change quite significantly. In particular, in the context of the non-flat PADE-II scenario, we see $w_1 = 1.60^{+0.16}_{-0.22}$ (at 68\% CL for SN+BAO+BBN+$H(z)$) is shifted to $0.231 \pm 0.077$ (at 68\% CL for SN+BAO+BBN+$H(z)$+GW). 
The $w_2$ parameter is found to be significantly affected in all three parametrizations after the inclusion of GWSS. We find that $w_2$ for PADE-I is shifted from $1.90 \pm 0.15$ (68\% CL, SN+BAO+BBN+$H(z)$) to $0.97 \pm 0.15$ (68\% CL, SN+BAO+BBN+$H(z)$+GW); for
SPADE-I, it is shifted from $0.67 \pm 0.59$ (68\% CL, SN+BAO+BBN+$H(z)$)  to $-0.12 \pm 0.17$ (68\% CL, SN+BAO+BBN+$H(z)$+GW); and for PADE-II, it is shifted from $-2.01^{+0.26}_{-0.20}$ (68\% CL, SN+BAO+BBN+$H(z)$) to $-0.903^{+0.092}_{-0.12}$ (68\% CL, SN+BAO+BBN+$H(z)$+GW). However, we note that for both PADE-I and PADE-II, $(w_1, w_2) \neq (0, 0)$ at more than 99\% CL with an improvement of $\Delta \chi^2 = -3.5$ (for PADE-I) and $\Delta \chi^2 = -3.3$ (for PADE-II), see Table~\ref{tab:IC}. While for SPADE-I, even though $w_2 =0$ is allowed within 68\% CL, but an improvement in the fit with respect to the non-flat $\Lambda$CDM model is observed with $\Delta \chi^2 = -2.4$.  We further observe that the matter density parameter also gets affected mildly after the inclusion of GWSS to SN+BAO+BBN+$H(z)$. In Table~\ref{tab:flat-GW} we present the constraints on the same scenarios but in a spatially flat universe from which we notice that the constraints on $w_1$, $w_2$ which are the key parameters in determining the dynamical nature of these parametrizations are changed but the dynamical nature of the PADE parametrizations are not compromised.

The effects of GWSS dataset are also visible when this is added to the combined dataset SN+BAO+BBN+$H(z)$+CMB. Our major observations in this case are as follows. The inclusion of GWSS to  SN+BAO+BBN+$H(z)$+CMB, changes the results on $\Omega_{k0}$ quite significantly. We find that  when the GWSS dataset is added to SN+BAO+BBN+$H(z)$+CMB, the possibility of the non-flat universe is pronounced for all three parametrizations in contrary to the results from SN+BAO+BBN+$H(z)$+CMB. In particular, in presence of GWSS with this combined dataset, we find $\Omega_{k0} \neq 0$ at more than 95\% CL for PADE-I and SPADE-I, and $\Omega_{k0} \neq 0$ at more than 68\% CL, for PADE-II.  
We further notice that in presence of GWSS, the phantom nature of $w_0$ obtained in all three scenarios is pronounced  and the uncertainties in $w_0$ are mildly reduced. Focusing on the $w_1$ parameter in PADE-I and SPADE-II, we find that the mean values and the uncertainties are changed after the addition of GWSS to SN+BAO+BBN+$H(z)$+CMB: for PADE-I, $w_1 = -0.480 \pm 0.029$ (68\% CL, SN+BAO+BBN+$H(z)$+CMB) is shifted to $w_1 = -0.397 \pm 0.053$ (68\% CL, SN+BAO+BBN+$H(z)$+CMB+GW); and for PADE-II, $w_1 = 1.89 \pm 0.29$ (68\% CL, SN+BAO+BBN+$H(z)$+CMB) gets shifted to $w_1 = -0.257 \pm 0.071$ (68\% CL, SN+BAO+BBN+$H(z)$+CMB+GW). For $w_2$, massive changes are observed in its mean values and  uncertainties for PADE-I and PADE-II. We find that for PADE-I, $w_2 = 1.81 \pm 0.10$ (68\% CL, SN+BAO+BBN+$H(z)$+CMB) is shifted to $w_2  = 1.464^{+0.072}_{-0.092}$ (68\% CL, SN+BAO+BBN+$H(z)$+CMB+GW) 
 and for PADE-II, $w_2 = -2.35 \pm 0.33$ (68\% CL, SN+BAO+BBN+$H(z)$+CMB) is shifted to $w_2 = -0.876 ^{+0.072}_{-0.092}$ (68\% CL, SN+BAO+BBN+$H(z)$+CMB+GW). 
Overall, we notice that $(w_1, w_2) \neq (0, 0)$ at many standard deviations  for PADE-I and PADE-II with an improvement of $\Delta \chi^2 = -2.2$ (for PADE-I) and $\Delta \chi^2 = -3.3$ (see Table~\ref{tab:IC}), and $w_2 \neq 0$ at more than 68\% CL for SPADE-I with an improvement of $\Delta \chi^2 = - 3.1$ (see Table~\ref{tab:IC}).  Hence, the evidence of dynamical DE is found within these parametrizations for this combined dataset.  This evidence does not alter in the absence of the curvature while the strength of evidence may alter depending on the constraints on the parameters, see Table~\ref{tab:flat-GW}.

Finally, for a better understanding on the effects of the GWSS on the cosmological parameters (in the non-flat universe), we have shown the whisker graphs of $\Omega_{k0}$ (Fig. \ref{fig:OMEGAK}) and the free parameters of the PADE parametrizations, namely, $w_0$ (Fig. \ref{fig:w0}), $w_1$ (Fig. \ref{fig:w1}) and $w_2$ (Fig. \ref{fig:w2}).  These whisker graphs clearly exhibit how GWSS dataset significantly affects the cosmological parameters. 
Thus, summarizing the effects, one can safely conclude that GWSS could be effective by providing the stringent constraints on the cosmological parameters subject to the availability of a large number of GW detections.

\subsection{Model comparison through various information criteria}
\label{sec-model-comparison}

Lastly, we assess the fitness of the PADE parametrizations with respect to all the combined datasets explored in this work. In order to do so 
we apply various information criteria, namely, the Akaike Information Criterion (AIC) \citep{Akaike:1974}, Bayesian Information Criterion (BIC) \citep{Schwarz:1974} and Deviance Information Criterion (DIC) \citep{spiegelhalter2002bayesian}. And to compare the ability of fitness of the models, one needs to select a reference model with respect to which the model comparison needs to be done. As $\Lambda$CDM cosmological model is undoubtedly one of the successful models, and we are investigating all the models in a non-flat universe, therefore, for our calculations, we choose the non-flat $\Lambda$CDM as the reference model. The AIC and BIC  are defined as follows 
\begin{eqnarray}
&&{\rm AIC} = \chi^2_{\rm min}+2K\;,\\
&&{\rm BIC} = \chi^2_{\rm min}+K\ln N\;,
\end{eqnarray}
where $K$ and $N$ are the number of free parameters and the total number of data points in the data combinations, respectively. Assuming these information criteria, given a set of cosmological models competing for the description of the same data samples, the preferred model is the one which has the minimum value of AIC and BIC. Hence, we can use the
differences $\Delta$AIC ($=$ AIC (model) $-$ AIC (non-flat $\Lambda$CDM)) and $\Delta$BIC ($=$ BIC (model) $-$ BIC (non-flat $\Lambda$CDM)).
The resulting $\Delta$AIC and $\Delta$BIC are used to determine the ``level of support for'' and ``evidence against'' each model respectively (see more details in \cite{Rezaei:2019xwo}).

The other criterion, DIC, employs both the Bayesian statistics and the information theory concepts \citep{spiegelhalter2002bayesian}. 
It is expressed as \citep{Liddle:2007fy}
\begin{eqnarray}
{\rm DIC} = D(\bar{{\bf p}})+2C_{\rm B}\;.
\end{eqnarray}
in which $C_{\rm B}=\overline{{D({\bf p})}}-D(\bar{{\bf p}})$ 
is the Bayesian complexity and over-lines imply the standard  mean value. 
Moreover, $D({\bf p})$ is the Bayesian deviation, which can be expressed as $D({\bf p})=\chi^2_{\rm t}({\bf p})$ in the case of exponential class of distributions [see more details in Refs. \cite{Trashorras:2016azl,Rezaei:2021qpq}]. In a similar fashion, we define $\Delta$DIC ($=$ DIC (model) $-$ DIC (non-flat $\Lambda$CDM)). In Table~\ref{tab:IC} we presented the results of model comparison obtained from different information criteria considering all the combined datasets.  
In the following we discuss the results of the information criteria. 

\begin{itemize}
\item {\bf AIC:} Assuming the results of this criterion (see Table~\ref{tab:IC}) we observed that for all the datasets considered here, non-flat $\Lambda$CDM remains  the best cosmological scenario. Let us focus on the results inferred from the combined datasets SN+BAO+BBN+$H(z)$ and SN+BAO+BBN+$H(z)$+CMB considering the usual cosmological probes. For the combined dataset SN+BAO+BBN+$H(z)$,  we find ``significant support'' for SPADE-I and ``considerably less support'' for PADE-I and PADE-II with respect to the non-flat $\Lambda$CDM model. However,  
the inclusion of CMB to SN+BAO+BBN+$H(z)$ (i.e. for the combined dataset SN+BAO+BBN+$H(z)$+CMB) worsens the fits for all the parametrizations where we have $\Delta {\rm AIC} > 4$  which means ``considerably less support'' for these models with respect to the non-flat $\Lambda$CDM. On the other hand, when the GWSS are added to these aforementioned datasets, we find that the values of $\Delta {\rm AIC}$ decrease for all the PADE parametrizations where for both SN+BAO+BBN+$H(z)$+GW and SN+BAO+BBN+$H(z)$+CMB+GW, ``significant support'' for SPADE-I is found  while ``less support'' is seen for PADE-I and PADE-II with respect to the non-flat $\Lambda$CDM model. At this point, it is important to note that even though $\chi^2$ values obtained for the PADE parametrizations are less than the $\chi^2$ values for the non-flat $\Lambda$CDM model 
for all the combined datasets (with or without GWSS) except in one case where $\chi^2_{min} = 1061.4$ (for SPADE-I) is greater than $\chi^2_{min} = 1061.3$ (non-flat $\Lambda$CDM) for SN+BAO+BBN+$H(z)$+CMB, however, AIC values (and consequently the $\Delta {\rm AIC}$ values) for the PADE parametrizations are increasing due to the large number of free parameters in these scenarios, and hence, the number of parameters are penalizing the fitness of these parametrizations.

\item {\bf BIC:} Our results from the current criterion are the same as those of AIC.  Considering all the data combinations (with and without GWSS), we find that $\Delta {\rm BIC} > 10$ for other all the PADE parametrizations which means ``very strong evidence'' against these scenarios. Hence, according to this criterion, non-flat $\Lambda$CDM remains as the best fitted cosmological model. 

\item {\bf DIC:} The results of DIC are a bit different from those of other criteria. Although, ${\rm DIC}$ values indicate that for all the data combinations (with GWSS and without GWSS), the non-flat $\Lambda$CDM remains the best fitted model, but, here we observe ``significant support'' for the DE scenarios described by PADE parametrizations which is quantified through $\Delta {\rm DIC}$. In contrast with AIC and BIC  which only penalize all the involved parameters, DIC penalizes just those parameters which contribute to the fit in an actual way. Assuming all of these aspects, DIC seems to be a more accurate tool for checking models with  more extra parameters compared to the concordance model [see the detailed discussions in \citep{Liddle:2007fy,Rezaei:2021qpq}].
\end{itemize}

\section{Summary and Conclusions}
\label{sec-conclusion}

The possibility of a non-zero curvature of the universe indicated by the recent observations \cite{Aghanim:2018eyx,Handley:2019tkm,DiValentino:2019qzk,DiValentino:2020hov,Efstathiou:2020wem} has been one of the central themes of modern cosmology at present moment and this is the main concern of this article. The inclusion of the curvature parameter in the analysis of DE and modified gravity models is a natural assumption and it provides a complete description of the underlying  cosmological model. As the sensitivity of the astronomical data is growing with time, therefore, the best way is to allow the curvature parameter into the analyses of the cosmological models and allow the observational data to decide its fate. In the present article we have considered some generalized DE parametrizations motivated from the PADE parametrization, labeled in this work as PADE-I, SPADE-I and PADE-II which were studied in Ref.~\cite{Rezaei:2017yyj} but in a curvature free universe. We initially constrained the underlying scenarios using the combined datasets from current cosmological probes, namely, SN+BAO+BBN+$H(z)$ and SN+BAO+BBN+$H(z)$+CMB and then we have included the 1000 mock GWSS luminosity distance measurements from the Einstein Telescope with these combined probes, (i.e. SN+BAO+BBN+$H(z)$ and SN+BAO+BBN+$H(z)$+CMB). The results on various model parameters of the proposed parametrizations are shown in The results are shown in 
Tables~\ref{tab:best}, \ref{tab:flat1}, \ref{table-H0}, \ref{tab:best-GW}, \ref{tab:flat-GW}. From the combined probes SN+BAO+BBN+$H(z)$ and SN+BAO+BBN+$H(z)$+CMB, 
we do not find any evidence of the curvature of our universe within the proposed parametrizations.  This result is in agreement with the earlier reports \cite{Efstathiou:2019mdh,Efstathiou:2020wem} where Planck in combination with some external probe, e.g. baryon acoustic oscillations prefers a flat universe. 
Concerning the dark energy properties, for both PADE-I and PADE-II, we find evidences of $(w_1, w_2) \neq (0, 0)$ at more than 99\% CL (see Figs. \ref{fig:padeI}, and \ref{fig:padeII}) for both the combined datasets, while for SPADE-I, although for SN+BAO+BBN+$H(z)$, we find an evidence of $w_2 \neq 0$ at more than 68\% CL, but in presence of CMB (i.e. for the combined dataset SN+BAO+BBN+$H(z)$+CMB), this evidence goes away. 
Thus, for both PADE-I and PADE-II, the evidence of dynamical DE, as a possible source of the current cosmological tensions as argued in Refs. \cite{Sola:2016jky,Zhao:2017cud,Zhang:2019jsu} is strongly supported at many standard deviations.

Quite interestingly, the inclusion of GWSS to both SN+BAO+BBN+$H(z)$ and SN+BAO+BBN+$H(z)$+CMB significantly affects most of the cosmological parameters. The whisker graphs in Figs. \ref{fig:OMEGAK}, \ref{fig:w0}, \ref{fig:w1}, \ref{fig:w2}
further strengthen the effects of GWSS on the key cosmological parameters. We notice that the evidence for a non-flat universe is pronounced for both SN+BAO+BBN+$H(z)$+GW and SN+BAO+BBN+$H(z)$+CMB+GW. This indication is promising for the cosmological data analysis in presence of the GWSS luminosity distance measurements and this suggests that the future cosmological probes could be very promising to offer new results on the curvature of our universe. In addition, for both PADE-I and PADE-II, concerning the nature of $w_1$ and $w_2$, we find them to be non-zero at more than 99\% CL, that means that they strongly indicate for a dynamical DE equation of state. While for SPADE-I, we notice that for SN+BAO+BBN+$H(z)$+GW, no strong evidence of $w_2 \neq 0$ is found, but for SN+BAO+BBN+$H(z)$+CMB+GW dataset,  a mild indication for $w_2 \neq 0$ is observed.

Finally, we assessed the ability of the PADE parametrizations in terms of their fitness with the current observational data by performing three well known information criteria, namely, AIC, BIC, DIC (see Table~\ref{tab:IC}). We find that for all the information criteria, the non-flat $\Lambda$CDM model remains the best choice compared to all the PADE parametrizations in a nonflat universe. Precisely, we find that BIC does not support any of the PADE parametrizations for {\bf any} datasets. In terms of AIC,  even though SPADE-I has a considerable support for SN+BAO+BBN+$H(z)$, but this support goes away when CMB data are included to SN+BAO+BBN+$H(z)$. However, interestingly, DIC shows considerable support towards all the PADE parametrizations for both SN+BAO+BBN+$H(z)$ and SN+BAO+BBN+$H(z)$+CMB. The support coming from DIC is stimulating for two reasons: DIC uses the whole sample space in contrast with AIC and BIC which use only the best fit likelihood and DIC penalizes those parameters which contribute to the fit in an actual way unlike AIC and BIC penalizing all the involved parameters.

Based on the outcomes of the present article, one can see that the PADE parametrizations are very appealing in the context of DE phenomenology. In particular, the evidence of dynamical DE at many standard deviations and the hints for a spatial curvature of the universe after the inclusion of future GWSS data clearly signal that the cosmological probes from upcoming astronomical  missions  will bring more exciting news on the physics of the dark sector of our universe and its curvature.

\section{Acknowledgments}
We thank the referee for his/her constructive comments and suggestions which helped us to improve the quality and presentation of the article significantly. 
SP acknowledges the financial support from  the Department of Science and Technology (DST), Govt. of India under the Scheme  ``Fund for Improvement of S\&T Infrastructure (FIST)'' [File No. SR/FST/MS-I/2019/41]. WY was supported by the National Natural Science Foundation of China under Grants No. 12175096 and No. 11705079, and Liaoning Revitalization Talents Program under Grant no. XLYC1907098. DFM thanks the Research Council of Norway for their support and the UNINETT Sigma2 -- the National Infrastructure for High Performance Computing and Data Storage in Norway.

\bibliography{biblio}

\end{document}